%% file: main.tex
\documentclass[sigconf, nonacm]{acmart}

\AtBeginDocument{%
  }

\copyrightyear{2026}
\acmYear{2026}
\setcopyright{cc}
\setcctype{by}

\input{dependencies.tex}

\begin{document}

\title{Identifying and Upweighting Power-Niche Users to Mitigate Popularity Bias in Recommendations}

\author{David Liu}
\affiliation{%
  \institution{Cornell University}
  \city{Ithaca}
  \state{NY}
  \country{USA}}
\email{davidliu@cornell.edu}
\orcid{0000-0002-2129-447X}

\author{Erik Weis}
\affiliation{%
  \institution{Northeastern University}
  \city{Boston}
  \state{MA}
  \country{USA}}
\email{weis.er@northeastern.edu}
\orcid{0009-0000-7275-8230}

\author{Moritz Laber}
\affiliation{%
  \institution{Northeastern University}
  \city{Boston}
  \state{MA}
  \country{USA}}
\email{laber.m@northeastern.edu}
\orcid{0000-0002-0273-584X}

\author{Tina Eliassi-Rad}
\affiliation{%
  \institution{Northeastern University}
  \city{Boston}
  \state{MA}
  \country{USA}}
\email{t.eliassirad@northeastern.edu}
\orcid{0000-0002-1892-1188}

\author{Brennan Klein}
\affiliation{%
  \institution{Northeastern University}
  \city{Boston}
  \state{MA}
  \country{USA}}
\email{b.klein@northeastern.edu}
\orcid{0000-0001-8326-5044}

\renewcommand{\shortauthors}{David Liu, Erik Weis, Moritz Laber, Tina Eliassi-Rad, \& Brennan Klein}

\begin{abstract}
Recommender systems have been shown to exhibit popularity bias by over-recommending popular items and under-recommending relevant niche items. We seek to understand niche users in benchmark recommendation datasets as a step toward mitigating popularity bias. We find that, compared to mainstream users, niche-preferring users exhibit a longer-tailed activity-level distribution, indicating the existence of users who both prefer niche items and exhibit high activity levels on platforms. We partition users along two axes: (1) activity level (``power'' vs. ``light'') and (2) item-popularity preference (``mainstream'' vs. ``niche''), and show that in three benchmark datasets, the number of power-niche users (high activity and niche preference) is statistically significantly larger than expected. We also find that interaction data from power-niche users is especially valuable for improving recommendations for not only niche but also mainstream users.  In contrast, many existing popularity bias mitigation methods have focused on upweighting niche users regardless of activity level. Motivated by the value of power-niche user data, we propose \texttt{PAIR} (Popularity-and-Activity-Informed Reweighting), a framework for reweighting the Bayesian Personalized Ranking (BPR) loss that simultaneously reweights based on user activity level and item popularity, upweighting power-niche users the most. We instantiate the framework on both deep and shallow collaborative filtering models, and experiments on benchmark datasets show that \texttt{PAIR} reduces popularity bias and can increase overall performance. Although existing popularity-bias mitigation methods yield a trade-off between performance and bias, our results suggest that considering both user activity level and popularity preference leads to Pareto-dominant performance.
\end{abstract}




\maketitle

\input{sections/1_intro}
\input{sections/2_related_work}
\input{sections/3_user_analysis}
\input{sections/4_data_value}
\input{sections/5_method}
\input{sections/6_experiments}
\input{sections/7_conclusion}

\begin{acks}
T.E.R. was supported in part by the Joseph E. Aoun Endowment.
D.L. was supported by the National Science Foundation Graduate Research Fellowship Program (GRFP) under Grant No. 2439018.
E.W. was supported by the National Science Foundation under Grant No. 2244340.
Any opinions, findings, and conclusions or recommendations expressed in this material are those of the author(s) and do not necessarily reflect the views of the National Science Foundation.
\end{acks}

\clearpage
\balance
\bibliographystyle{ACM-Reference-Format}
\bibliography{references}

\clearpage
\input{sections/appendix}

\end{document}

%% file: dependencies.tex
\usepackage{multirow}
\usepackage{subcaption}
\usepackage{natbib}
\usepackage{hyperref}
\usepackage{enumitem}

\usepackage{graphicx,booktabs,tabularx,multirow,makecell}
\usepackage[table]{xcolor} % for row colors
\usepackage{adjustbox}

\definecolor{bestrow}{RGB}{240,244,248} % very light blue-gray
\newcommand{\bestrow}{\rowcolor{bestrow}}      % highlight whole rows

%% file: sections/1_intro.tex
\section{Introduction} \label{sec:intro}

Recommender systems are known to pervasively exhibit popularity bias, disproportionately recommending already popular items and perpetuating popularity feedback loops \cite{klimashevskaia2024survey, steck2011item, abdollahpouri2021user}. Popularity bias harms both users and items. Users are harmed because those who have niche preferences unfairly receive lower-quality recommendations \cite{abdollahpouri2021user, abdollahpouri2020connection}. And, items are harmed because popularity bias limits the visibility of new content creators \cite{ionescu2023group}, leading to monopolies for the already popular. In response, a host of algorithms have been proposed to mitigate popularity bias, with a focus on regularizing and reweighting loss functions \cite{rendle2009bpr, rhee2022countering, jannach2015what} and post-process re-ranking \cite{abdollahpouri2021user, zehlike2017fair}. Field studies have shown that mitigation techniques do not necessarily hinder user engagement \cite{klimashevskaia2023evaluating} and can even increase the user's sense of discovery \cite{ungruh2024putting}.

Although many algorithms have been proposed to mitigate popularity bias, the question of \emph{who} consumes niche items remains largely unexplored. We ask the following research questions:
\begin{itemize}[leftmargin=*, topsep=2pt, itemsep=1pt, parsep=0pt]
    \item \textbf{RQ1:} Beyond item preference, how do the behaviors of niche users differ from mainstream users in large-scale recommendation datasets?
    \item \textbf{RQ2:} How can we leverage these differences in behavior to mitigate popularity bias?
\end{itemize}
We answer these questions through the following three contributions. The first contribution answers RQ1, while the latter two answer RQ2.

\textbf{Motivating Data Analysis (RQ1).} In Sec. \ref{sec:data-analysis} and \ref{sec:data-value}, we begin with a motivating data analysis of benchmark recommendation datasets and divide users into four quadrants based on their (1) activity level and (2) item-popularity preference. We find that there exists a statistically significant group of high-activity, niche-preferring users, which we term \emph{power-niche users}. In contrast, among mainstream users, there is a significant group of \emph{light-activity} users.
Next, we assess the value of interaction data from each of the four user quadrants to improve recommendations for niche users. While previous bias mitigation work has focused on upweighting interactions from all niche users, we find that compared to light-niche users, data from power-niche users are more valuable for improving both niche and overall recommendation performance. In the process, we propose a method for valuing user groups in recommender systems and transductive machine learning models, more generally.

\textbf{Reweighting Framework (RQ2).} In Sec.~\ref{sec:method}, guided by the promising value of power-niche user data, we design a framework, Popularity-and-Activity-Informed Reweighting (\texttt{PAIR}), that simultaneously reweights for user activity level and item popularity. Our framework adapts the Bayesian Personalized Ranking (BPR) \cite{rendle2009bpr} loss function and introduces two parameters: one for controlling the weight of power users and another for the weight of niche items. We optimize the parameters through validation-set grid search. Though past works have focused on reweighting for item popularity, either in static \cite{zhao2013opinion, steck2011item} or longitudinal \cite{jannach2015what, zhu2021popularitybias, ferraro2020exploring} settings, to the best of our knowledge, we are the first to consider both user activity level and item popularity heterogeneity for popularity bias. Our framework allows us to give greater weight to power-niche users.

\textbf{Evaluation (RQ2).} In Sec. \ref{sec:eval}, our experiments show that by upweighting power-niche users in the BPR loss function, we can \emph{increase} recommendation performance and decrease popularity bias. Further, we show that existing reweighting and regularization approaches reduce popularity bias at the cost of performance, while our approach (\texttt{PAIR}) yields Pareto-dominant performance by accounting for user-activity level.

%% file: sections/2_related_work.tex
\section{Related Works}\label{sec:related-works}
We review the extensive literature on measuring and mitigating popularity bias. 
We also review the literature on assigning value to user data, which is related to our motivating analysis regarding the value of user quadrants.

\subsection{Popularity Bias}
The popularity-bias literature has proposed numerous bias metrics and methods for mitigating this bias. Most metrics are centered on the items. Common metrics include summary statistics such as the coverage of the recommended items over the item set \cite{adomavicius2012improving} or the average popularity of recommended items \cite{abdollahpouri2019managing}. Popularity bias has also been defined in terms of equal opportunity, ensuring that head (popular) and tail (less popular) items have balanced true-positive rates \cite{zhu2021popularitybias}. From the user perspective, \citet{abdollahpouri2021user} define popularity bias in terms of popularity calibration, that is, ensuring that the distribution of recommended item popularities for a user matches the historical popularity distribution. In our evaluation in Sec. \ref{sec:eval}, we evaluate bias mitigation with a mixture of item and user-centric metrics.
In terms of mitigating popularity bias, past approaches can be categorized as either regularization  \cite{abdollahpouri2017controlling, zhu2021popularityopportunity}, reweighting \cite{steck2011item, zhao2013opinion,schnabel2016recommendations,lee2021dual}, or post-processing \cite{zhu2021popularityopportunity,abdollahpouri2021user,abdollahpouri2019managing,antikacioglu2017post,zehlike2017fair}. Our approach is most similar to Inverse Propensity Weighting (IPW) reweighting approaches, as discussed further in Sec. \ref{sec:prior-work}, but in addition to item propensity, we consider user activity level. 

While prior work on popularity bias has considered item popularity alongside user activity level \cite{rahmani2022experiments, naghiaei2022CPFair}, these studies do not define power-niche users or quantify their value. For example, \citet{abdollahpouri2019unfairness} identify a negative correlation between user activity level and item-popularity preference, suggesting the existence of high-activity niche users. Our work shows that not only do power-niche users exist, but the number of power-niche users is statistically significantly greater than expected across multiple datasets. Alternatively, while algorithmic approaches to mitigating popularity bias often frame a bias-performance trade-off \cite{abdollahpouri2017controlling, zanon2022balancing}, our work shows that upweighting power-niche users can decrease bias while increasing performance.

\subsection{Value of User Data}

Quantifying the value of data points has a number of uses in machine learning, such as ascribing economic value to observations \cite{henderson2023} or providing efficient guides for the collection of new data \cite{ghorbani2019data}.
A baseline approach for such a task is the Leave-One-Out (LOO) technique, where the value of a point is the marginal increase in a model's performance score when the point is added to the remaining data \cite{cook1997}.
Motivated by the theory of cooperative games, Data Shapley \cite{ghorbani2019data} expands on this idea with a score that evaluates a data point on its marginal value when added to a random subset of the data.
Despite the prominence of Data Shapley, other promising alternatives exist \cite{Yan_Procaccia_2021}, leaving the problem of data valuation somewhat unsettled.

The problem becomes harder still when data points have relational dependencies, as opposed to being drawn i.i.d. from an underlying distribution.
Recommender systems are one such example.
In this setting, defining a single score based on data subsets raises a number of questions, such as whether nodes or edges constitute the appropriate unit of measurement and whether one should be concerned with the potentially heterogeneous contributions to model performance a data point may have through its relational dependencies.
One possible treatment of the graph data valuation problem is to limit its scope to local and conditional measures of data value, an approach which comes from the literature on GNN explainability \cite{chi2025precedenceconstrained, graphSVX, gnnshap}.
For instance, \citet{chi2025precedenceconstrained} develop PC-Winter, a score for the marginal value of a node $i$ in the context of another node $j$'s neighborhood.
Although such values can be averaged at the node or edge level to produce a single score, the resulting value still lacks a clear interpretation in our context, where we are interested in comparing globally different compositions of the data by users and items. 
Instead, our analysis is closer in spirit to LOO measures, but performed at the level of groups of users as opposed to single data points.
We choose nodes (as opposed to edges), since this is the frame most likely to guide the collection of new data in recommender systems.

%% file: sections/3_user_analysis.tex
\section{Understanding Niche Users} \label{sec:data-analysis}
In this section, we conduct a motivating data analysis of popular benchmark datasets to better understand who engages with niche items and what other behaviors distinguish niche users from others. We examine the relationship between user activity level and item preference and find that there is a statistically significant number of users who exhibit a high activity level and preference for niche items. We call this group \emph{power-niche users}. In contrast, among users with mainstream preferences, there are statistically significantly fewer power users. Our notation is summarized in Appendix \ref{sec:notation}.

We define niche users as follows. Given a binary interaction matrix $\mathbf{D} \in \{0, 1\}^{n\times m}$ among $n$ users and $m$ items, where $\mathbf{D}_{uj} = 1$ indicates user $u$ interacted with item $j$, let the popularity of item $j$, $d_j$, be the number of users that have interacted with the item, i.e. $d_j = \sum_u \mathbf{D}_{uj}$. Similarly, let the activity level of user $u$, $d_u$, be the number of interactions for user $u$, i.e. $d_u = \sum_j \mathbf{D}_{uj}$. We then define a user's ``item-popularity preference'' as the average popularity of all items the user has interacted with. Larger values suggest that the user is mainstream, and lower values suggest the user is niche. In terms of $\mathbf{D}$, the item-popularity-preference of user $u$ is
\begin{equation}\label{eqn:item-pop-pref}
    \text{Item Pop. Pref. for user } u=\frac{\sum_{j=1}^m \mathbf{D}_{uj} d_j}{d_u}.
\end{equation}
We label a user as a niche user if their item-popularity preference is below the median item-popularity preference among all users. 

Our motivating analysis is based on the Gowalla, Yelp2018, and Amazon-Book binary implicit-feedback benchmark datasets \cite{wang2019neural,lee2024revisiting}, which are detailed in Table \ref{tab:datasets}.
\begin{table}[ht]
    \centering
    \caption{Benchmark datasets used in this paper}
    \begin{tabular}{llrr}
         \toprule
         \textbf{Dataset} & \textbf{Type} & \textbf{Users} & \textbf{Items}\\
         \midrule
         Gowalla & Location check-ins & 29,858 & 40,981 \\
         Yelp2018 & Reviews of businesses & 31,668 & 38,048 \\
         Amazon-Book & Reviews of books & 52,643 & 91,599\\
         \bottomrule
    \end{tabular}
    \label{tab:datasets}
\end{table}

\subsection{Identifying Power-Niche Users}
\begin{figure}[ht]
    \centering
    \includegraphics[width=\linewidth]{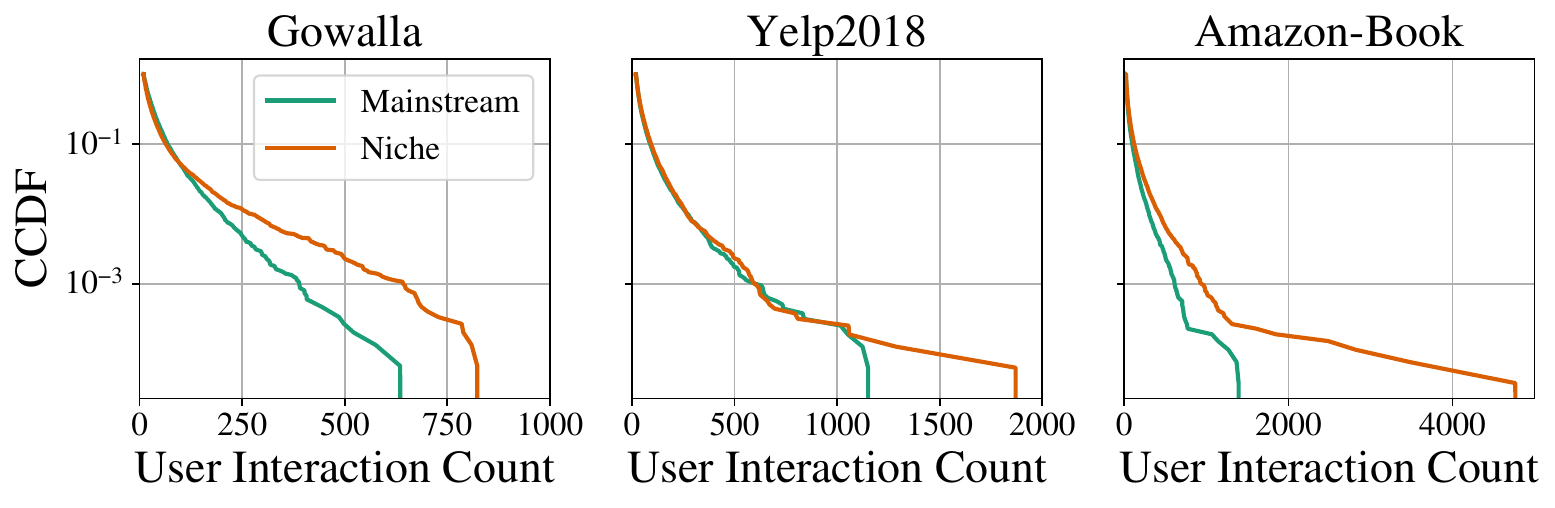}
    \Description[User activity level CCDF plots]{CCDF plots of user interaction counts for the Gowalla, Yelp, and Amazon datasets for niche and mainstream users. For each dataset, there is an activity-level distribution for mainstream users and a distribution for niche users.}
    \caption{Distribution of user activity levels for mainstream and niche users in the Gowalla, Yelp2018, and Amazon-Book datasets. Niche users have below-median item-popularity preference.
    In each dataset, the distribution of activity levels for niche users is longer (further right) than the distribution for mainstream users, indicating the existence of users who prefer niche items and exhibit high activity levels.}
    \label{fig:activity-level-distributions}
\end{figure}

\paragraph{Longer activity-level tail for niche users.}
In addition to examining the item preferences of users, we consider their activity levels, or the number of observed interactions. Let us define two groups of users: niche and mainstream. Niche users are those with below-median item-popularity preference, as defined in Eq. \eqref{eqn:item-pop-pref}, and mainstream users are all other users. Fig. \ref{fig:activity-level-distributions} visualizes the activity-level distributions for the two groups of users, where a point $x, y$ on a curve indicates that a $y$ fraction of users in the group have at least $x$ interactions. The plots show that compared to mainstream users, the distributions for niche users have a longer tail, indicating the presence of high-activity niche-preferring users.

\paragraph{Quadrants of users.}
We partition the users in each dataset into four quadrants. The first axis is activity level, where users are either above (``power'') or below (``light'') the median. The second axis is item-popularity preference, where, again, users are either above (``mainstream'') or below (``niche'') the median.

To contextualize the user quadrants, in Table \ref{tab:quadrant-summary}, we provide summary statistics for the quadrants across the three datasets. Although the quadrants have a similar number of users, power-niche users account for $38.5$-$44.5\%$ of all observed interactions (``Inter.''). The final columns in Table \ref{tab:quadrant-summary} show that the average power-niche user, relative to all users, interacts with items with $42.3$-$51.5\%$ fewer interactions. Light-niche users are marginally more niche than power-niche users but yield far fewer interactions.

\input{sections/tables-and-figures/quadrant_summary_table}

\subsection{Significance of Power-Niche Users}
\begin{figure}[ht]
    \centering
    \includegraphics[width=\linewidth]{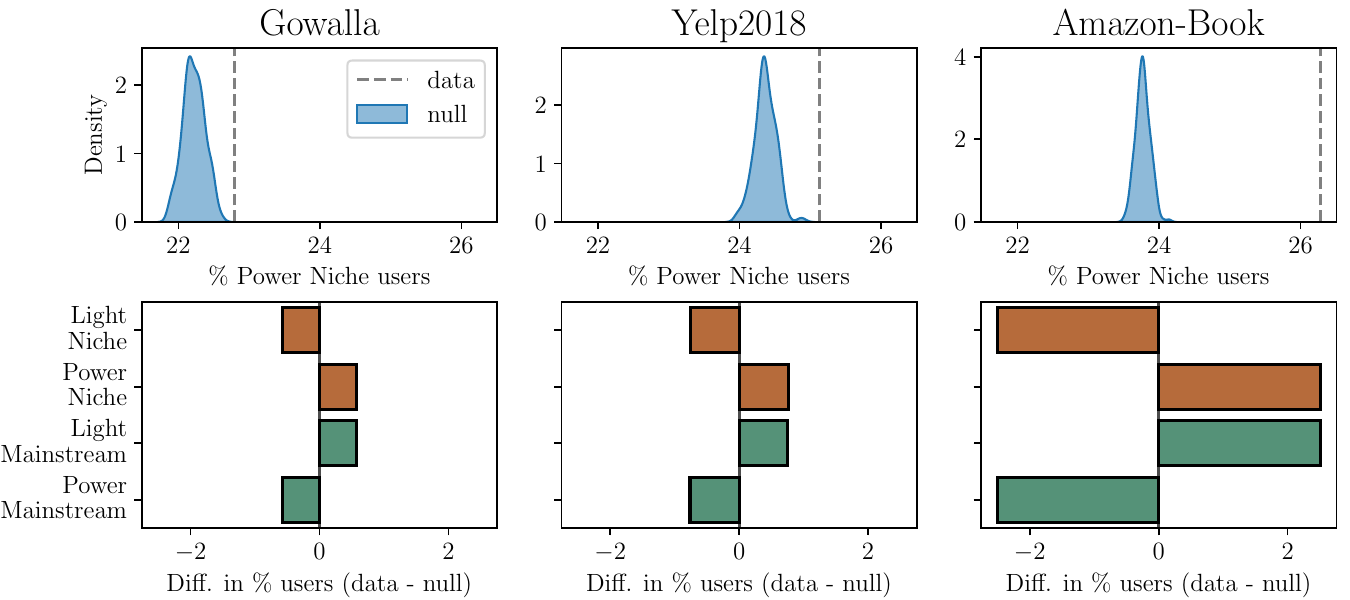}
    \Description[Statistical significance heat maps for user bins]{For each dataset, there is a binned heat map where bins partition users by interaction count quantile and item-popularity preference quantile. There is a collection of red bins in the bottom right of each figure, indicating a statistically significant increase over the null model.}
    \caption{Item interaction patterns of different user groups, compared to a configuration null model.
    Users are grouped by activity-level and item-popularity quantiles.
    Top row: we show the distribution of power-niche representation (expressed as a percentage) under the null model, as well as the same quantity observed in the data.
    Bottom row: we show the relative difference in percentage representation for the observed data and the null model for all user groups.
    For each dataset, the number of power-niche users---high activity level and niche item popularity preference---is higher than expected under the configuration null model.}
    \label{fig:configuration-model}
\end{figure}

To assess the statistical significance of the number of power-niche users we observe, we compare against a null configuration model \cite{newman2018configuration,fosdickConfiguringRandomGraph2018}. In the context of a user-item binary interaction graph, the null configuration model characterizes the space of all graphs with the same user activity levels and item popularities. To sample from this null model, we take the input interaction dataset and randomly rewire pairs of interactions: user-item pairs $(u_1, i_1)$ and $(u_2, i_2)$ become $(u_1, i_2)$ and $(u_2, i_1)$. For each null sample, we perform $10 E$ rewirings of the original interaction graph, where $E$ is the total number of user interactions. Fig. \ref{fig:configuration-model} illustrates the deviation in user counts in the observed datasets compared to samples from the null model.
We bin users into quadrants, and show the full distribution for the expected percent of users that belong to the power-niche category. In the three datasets, the observed number of power-niche users is over two standard deviations, our definition of statistical significance, greater than the mean under the null model.
Additionally, we show for all four categories the difference between each group's percent representation and its expected value under the null model. For mainstream users, the pattern is flipped, and there are fewer power users than expected.  

%% file: sections/tables-and-figures/quadrant_summary_table.tex
\begin{table}[t]
\centering
\caption{Summary statistics for the four user quadrants across the three benchmark datasets. The quadrants are comparably sized in terms of the number of users. The power-niche quadrant accounts for $38$-$45\%$ of all interactions. $\Delta$ Item-Pop. refers to the change in the average item-popularity preference among users in the quadrant relative to the average among all users. Power-niche users, as expected, prefer more niche items; however, they are marginally less niche than light-niche users.}

\resizebox{\columnwidth}{!}{%
\begin{tabular}{llrrr}
\toprule
\textbf{Quadrant} & \textbf{Dataset} & \textbf{Users (\%)} & \textbf{Inter. (\%)} & \textbf{$\Delta$ Item-Pop. (\%)} \\
\midrule
Light-Mainstream & Gowalla & 23.1\% & 9.6\% & +74.7\% \\
 & Yelp2018 & 25.6\% & 12.8\% & +50.6\% \\
 & Amazon-Book & 26.6\% & 11.7\% & +52.3\% \\
\addlinespace
Light-Niche & Gowalla & 27.2\% & 11.0\% & -58.3\% \\
 & Yelp2018 & 24.9\% & 12.5\% & -43.2\% \\
 & Amazon-Book & 23.7\% & 10.5\% & -48.3\% \\
\addlinespace
Power-Mainstream & Gowalla & 26.9\% & 40.8\% & +38.4\% \\
 & Yelp2018 & 24.4\% & 36.2\% & +34.6\% \\
 & Amazon-Book & 23.4\% & 33.3\% & +42.6\% \\
\hline
\addlinespace
Power-Niche & Gowalla & 22.8\% & 38.7\% & -51.5\% \\
 & Yelp2018 & 25.1\% & 38.5\% & -42.3\% \\
 & Amazon-Book & 26.3\% & 44.5\% & -47.2\% \\
\bottomrule
\end{tabular}%
}

\label{tab:quadrant-summary}
\end{table}

%% file: sections/4_data_value.tex
\section{Value of User Data in Recommender Systems}\label{sec:data-value}
In this section, we examine the value of user data from each of the user quadrants to improve overall and niche recommendations.
While we expected additional data from niche users to benefit niche users at the cost of mainstream users, we find that data from power-niche users is valuable for not just niche users but all users on average.
In Sec. \ref{sec:data-value-methodology}, we introduce our method for measuring the value of user quadrants, and in Sec. \ref{sec:power-niche-data-value}, we show that power-niche user data is valuable to both niche and mainstream users. 

\subsection{Quadrant Valuation Methodology}\label{sec:data-value-methodology}
Our method for evaluating the value of data from each user quadrant is similar to the Leave-One-Out (LOO) method but adapted for the user group setting. Given an input binary interaction dataset $\mathbf{D}\in\{0,1\}^{n\times m}$, we hold out $20\%$ of each user's interactions as a test set, $\mathbf{Y}$, and the remaining interactions are the training set $\mathbf{X}$, where $\mathbf{D} = \mathbf{X} + \mathbf{Y}$. 

Among the training dataset, we construct a fixed set of users representing a uniform random sample over the user base, such that, in expectation, the relative sizes of quadrants are preserved in the fixed set. Let $\mathcal{U}_{\text{fixed}}$ denote the set of fixed users. The purpose of our data valuation methodology is to determine the value of user data from each quadrant to improve the recommendations for $\mathcal{U}_{\text{fixed}}$. Let us partition all other users into their respective quadrants and denote these four sets as $\mathcal{U}_{\text{Light-Niche}}$, $\mathcal{U}_{\text{Light-Mainstream}}$, $\mathcal{U}_{\text{Power-Niche}}$, $\mathcal{U}_{\text{Power-Mainstream}}$. Crucially, these user groups are mutually exclusive from the fixed set. Here, we use the same user quadrant definition introduced in Sec. \ref{sec:data-analysis}.

In one trial of our data valuation methodology, we train two models. The first (the ``control'' model) has access only to training interactions from $\mathcal{U}_{\text{fixed}}$, while the second (the ``treatment'' model) has access to the same interactions plus a sample of users from a specified user quadrant. We then evaluate the difference in performance between these two models \emph{when evaluated on} $\mathcal{U}_{\text{fixed}}$. More formally, let
$f$ denote a prediction model;
let $X[\mathcal{U}_q]$ denote the rows of $\mathbf{X}$ corresponding to the set of users $q$;
let $R_f(\mathbf{X}, \mathbf{Y})$ denote the evaluation score (e.g. Recall@$k$) of $f$ trained on $\mathbf{X}$ and tested on $\mathbf{Y}$;
and let $\mathcal{S}$ denote a uniform-sampling function.
Then, the value of a quadrant $q$ is:
\begin{align}\label{eqn:data-value}
    \begin{split}
    \Delta_q &=
    R_f\left(\mathbf{X}[\mathcal{U}_{\text{fixed}} \cup \mathcal{S}\left(\mathcal{U}_q\right)], \mathbf{Y}[\mathcal{U}_{\text{fixed}}]\right) \\
    &- R_f\left(\mathbf{X}[\mathcal{U}_{\text{fixed}}], \mathbf{Y}[\mathcal{U}_{\text{fixed}}]\right) 
    \end{split}
\end{align}
In Eq. \eqref{eqn:data-value}, the crucial difference between the two recall terms is the difference in the indices for the training sets. One model is trained on $\mathbf{X}[\mathcal{U}_{\text{fixed}} \cup \mathcal{S}\left(\mathcal{U}_q\right)]$ and another is trained on $\mathbf{X}[\mathcal{U}_{\text{fixed}}]$. The test set is the same for both models.

\begin{figure}
    \centering
    \includegraphics[width=\columnwidth]{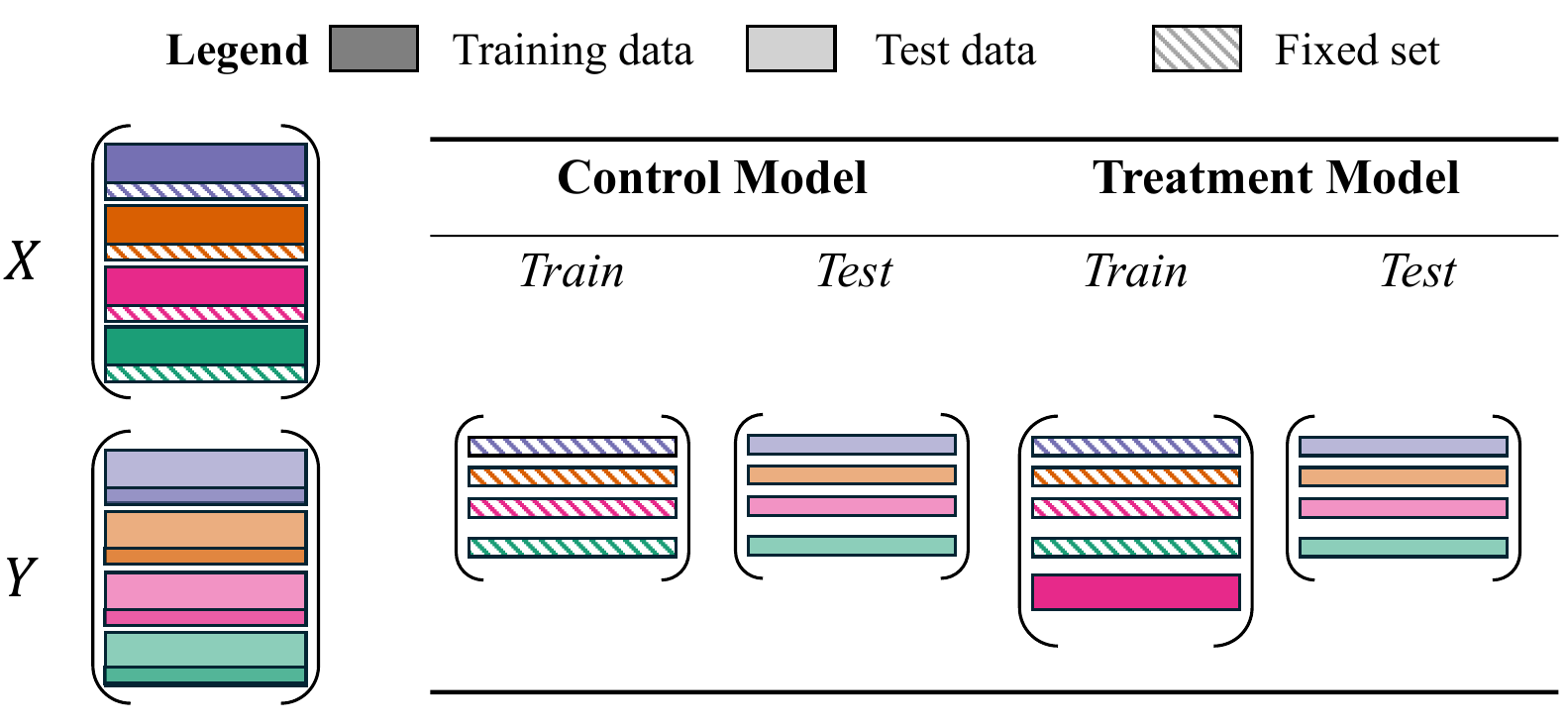}
    \Description[Quadrant methodology schematic]{Schematic of the data valuation procedure. A fixed set of users defines the evaluation set. A control model is trained using only fixed-set data, while a treatment model additionally includes sampled users from one quadrant. Both models are evaluated on the fixed users to estimate the marginal value of quadrant-specific data.}
    \caption{An overview of our method for valuing data from each user quadrant. A group of users is uniformly sampled at random and labeled the ``fixed set''. Two models are then trained: one using only training data from the fixed-set users and another that includes an additional sample from the target quadrant. The pink box under the treatment training data is representative of a sample---mutually exclusive of the fixed set---of power-niche users. Both models are evaluated on the fixed-set users. The procedure is repeated for multiple trials, and the value of a quadrant is the average change in performance. We evaluate the value of quadrant data at multiple treatment ratios, which is the relative size of the additional treatment data relative to the fixed set.}
    \label{fig:data-value-methodology}
\end{figure}

For a given quadrant, we average $\Delta_q$ over repeated samples of  $\mathcal{S}\left(\mathcal{U}_q\right)$ to yield a final score $\bar{\Delta}_q$ for the value of quadrant $q$. Fig. \ref{fig:data-value-methodology} summarizes our data valuation methodology, where the additional training data for the ``treatment'' model corresponds to $\mathbf{X}[\mathcal{S}\left(\mathcal{U}_q\right)]$.

\subsection{Value of Power-Niche Users for Improving Niche and Overall Performance}\label{sec:power-niche-data-value} 

Fig. \ref{fig:data-value} plots the value of each quadrant for two evaluation metrics: Recall@$k$ and Niche Recall@$k$, defined as the average Recall@$k$ computed only over niche users. For Fig. \ref{fig:data-value}, the number of users in the fixed set is $20\%$ of all users, and the treatment ratio on the x-axis refers to the number of additional users added in the treatment model relative to the fixed set. So, a treatment ratio of $0.5$ means that the additional quadrant sample $S\left(\mathcal{U}_q\right)$ has half as many users as the fixed set. As a baseline, we also report the data value of an additional random sample of users, ones that are not specific to any single quadrant, as shown in the black curves. The prediction model is LightGCN~\cite{he2020lightgcn}.

For Niche Recall, power-niche users are the most valuable for both Gowalla and Yelp2018 (increasing Niche Recall by up to $10\%$) and are the second most valuable for Amazon-Book. For overall Recall@$k$, power-niche users are consistently the most valuable across all three datasets. In contrast, light-mainstream users are often the least valuable. These results show that data from power-niche users is valuable for improving recommendations for both niche users, thus valuable for mitigating popularity bias, and for the user base as a whole.

For Yelp2018, adding additional training users, regardless of quadrant, only marginally improves overall performance. We hypothesize that for the Yelp2018 business-rating dataset, users are more idiosyncratic. This result also highlights the nontrivial nature of assessing data value in a transductive setting.

While the results in Fig. \ref{fig:data-value} correspond to a single fixed set $\mathcal{U_{\text{fixed}}}$, Appendix \ref{sec:fixed-set-robustness} shows that the value of power-niche users is robust to alternative fixed set samples.

\input{sections/tables-and-figures/data-value}

%% file: sections/tables-and-figures/data-value.tex
\begin{figure}[ht]
    \centering
    \begin{subfigure}{\columnwidth}
        \centering
        \includegraphics[width=\columnwidth]{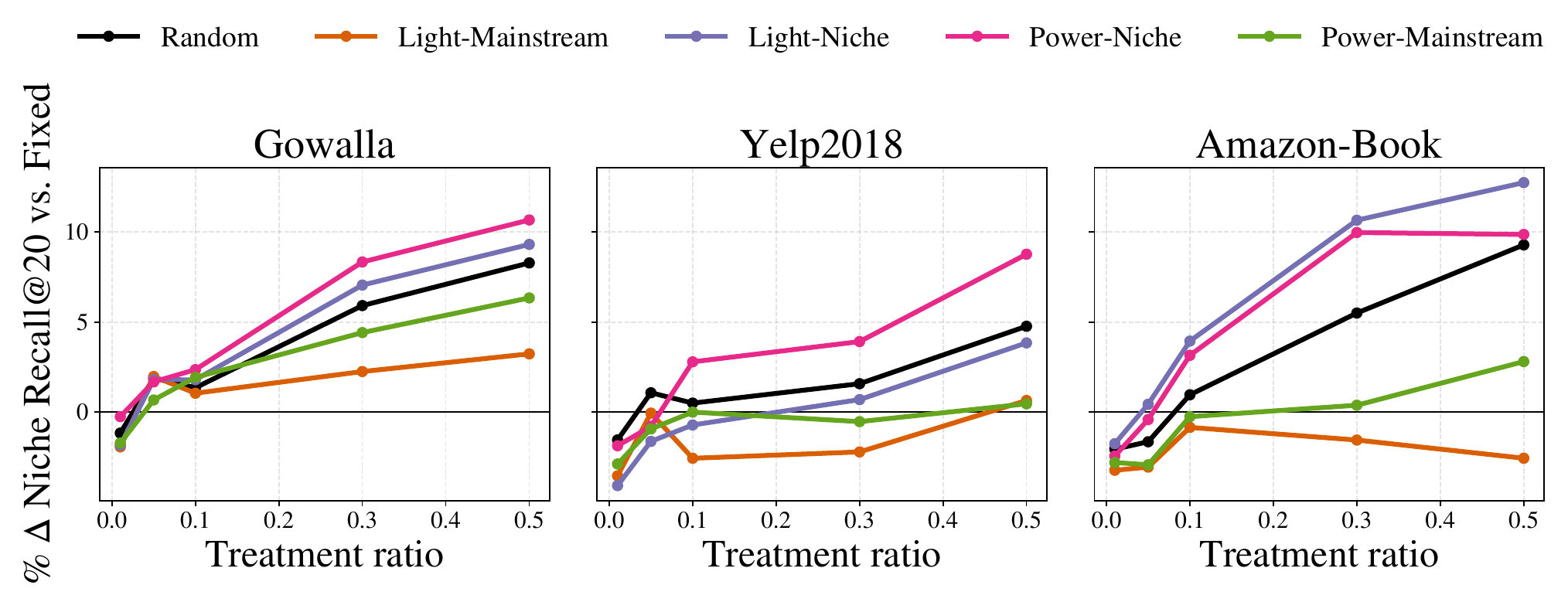}
    \end{subfigure}
    \begin{subfigure}{\columnwidth}
        \centering
        \includegraphics[width=\columnwidth]{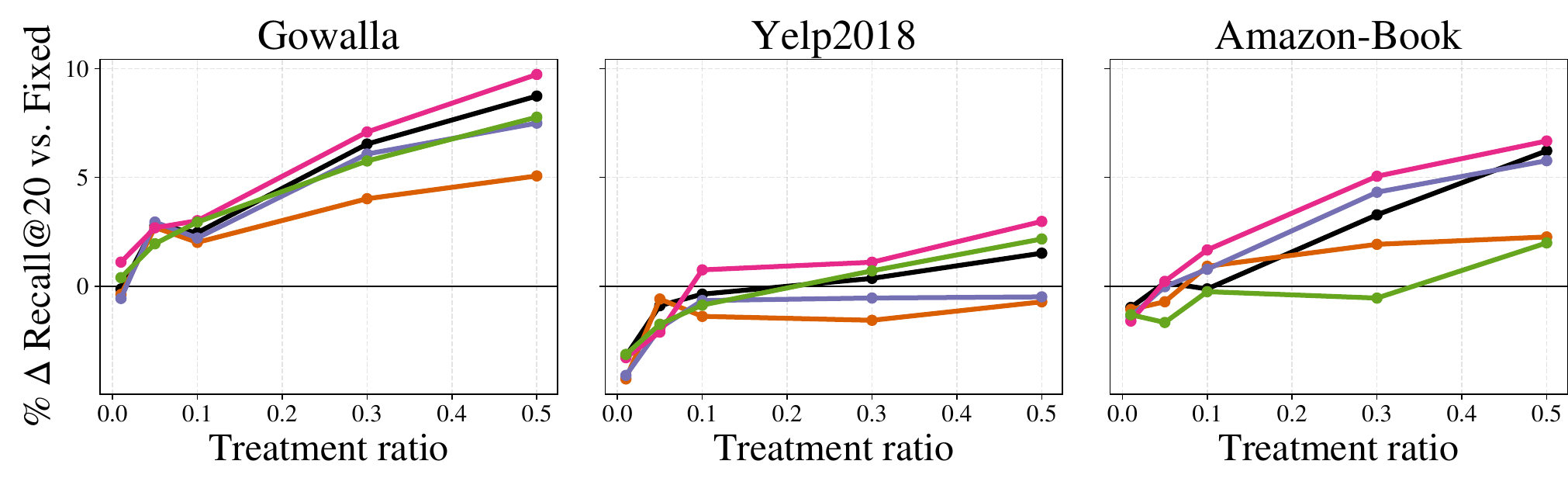}
    \end{subfigure}
    \Description[Data value experiment]{Line plots showing the relative change in Niche Recall@20 and overall Recall@20 as additional users from each quadrant are added to the training data. Results are shown for three datasets as a function of the treatment ratio, with power-niche users providing the largest gains in most settings.}
    \caption{To inform our reweighting, we examine the value of each quadrant for improving niche recommendation performance (top) and overall (bottom) performance. We measure the value of a quadrant by the relative change in performance (Recall@$k$) for all users and niche users when the model is re-trained after users from a specified quadrant are added to the training dataset. The test set is fixed. Power-niche users are not only valuable for improving niche user performance, but are also the most valuable quadrant for improving overall performance. }
    \label{fig:data-value}
\end{figure}

%% file: sections/5_method.tex
\section{PAIR: A Reweighting Framework for Bayesian Personalized Ranking} \label{sec:method}

We present Popularity-and-Activity Informed Reweighting (\texttt{PAIR}), a framework for simultaneously tuning the weights of niche items and high-activity users. Our motivating analyses identify that, in three benchmark datasets, there exists a statistically significant number of \emph{power-niche} users, who are valuable for improving niche and overall performance. \texttt{PAIR} is designed to increase the significance of these users in the Bayesian Personalized Ranking (BPR) loss function to mitigate popularity bias. 

\subsection{Our Framework} \label{sec:framework}
We consider a recommendation setting over a set of $n$ users, $\mathcal{U}$, and $m$ items, $\mathcal{I}$. There exists a set of ground-truth  interaction labels $\mathbf{Y} \in \{0, 1\}^{n \times m}$ where $\mathbf{Y}_{ui} = 1$ if and only if user $u$ interacts with item $i$. Let $l\left(\mathbf{Y}, \mathbf{\hat{Y}}\right)$ be a loss function for evaluating a set of user-item prediction scores $\mathbf{\hat{Y}} \in \mathbb{R}^{n \times m}$.

\texttt{PAIR} adapts the BPR loss function, which induces predictions where positive (interacted) user-item pairs have higher prediction scores than negative pairs. Specifically, the BPR loss function is 
\begin{equation}\label{eqn:bpr}
    l^{\text{BPR}}\left(\mathbf{Y}, \mathbf{\hat{Y}}\right)
    =
    -\sum_{u=1}^n \sum_{i\in \mathcal{N}_u} \sum_{j \notin \mathcal{N}_u} \ln \sigma\left(\mathbf{\hat{Y}}_{ui} - \mathbf{\hat{Y}}_{uj}\right).
\end{equation}
In Eq. \eqref{eqn:bpr}, $\mathcal{N}_u = \{i \mid \mathbf{Y}_{ui} = 1\}$ is the set of items that user $u$ has interacted with, and $\sigma$ is the sigmoid function.

To upweight the significance of power users and niche items, \texttt{PAIR} adapts the BPR loss function and optimizes an adapted loss function involving hyperparameters $\alpha, \beta$: 
\begin{equation} \label{eqn:our-loss}
     l^{\text{PAIR}}_{\alpha, \beta}\left(\mathbf{Y}, \mathbf{\hat{Y}}\right) = -\sum_{u=1}^n {d_u}^\alpha \left(\mathbb{E}_{\substack{i\sim \textrm{U}\left(\mathcal{N}_u\right) \\ j \sim \textrm{U}\left(\mathcal{I} \backslash \mathcal{N}_u\right)}}\left[{d_i}^{-\beta} \ln \sigma\left(\mathbf{\hat{Y}}_{ui} - \mathbf{\hat{Y}}_{uj}\right)\right]\right)
\end{equation}
$\alpha \in [0,1]$ and $\beta \in [0, \infty)$ are reweighting hyperparameters. Setting $\alpha=1,\,\beta=0$ is most closely aligned with a sample-based approximation of the BPR loss (see Eq.~\eqref{eqn:bpr}), whereas setting $\alpha=0,\,\beta=0$ mirrors the loss function of LightGCN. Our loss function flexibly interpolates, using $\alpha$, between these two extremes for taking user activity into account and, in addition, introduces via $\beta$ a way to take into account item popularity.
While we defined threshold-based quadrant boundaries for our motivating analysis, \texttt{PAIR} does not rely on group partitions and instead continuously upweights based on user activity level and item popularity.
Next, we discuss our reweighting rationale. 

\paragraph{Significance of Item Popularity.} Compared to the BPR loss function, we introduce a factor ${d_i}^{-\beta}$. Since $d_i$ is the popularity of item $i$ and $\beta >0$, the factor increases as the popularity of items decreases. Intuitively, the factor suggests that the loss should give greater weight to observed interactions with niche items because a user may interact with a popular item simply because popular items have greater exposure \cite{schnabel2016recommendations}. 

\paragraph{Significance of User Activity Level.} To adjust the significance of user activity level, we tune the number of positive items that we sample with replacement. Specifically, compared to the BPR loss, we do not sum over all positive and negative pairs $u, i$ and $u, j$ but rather take the expectation over a uniform sample of $i, j$ and scale the expectation by a factor ${d_u}^\alpha$. To optimize the scaled expectation, we sample 
$S_u = \texttt{round}\left(\frac{{d_u}^\alpha}{\sum_{u'\in \mathcal{U}} {d_{u'}}^\alpha} \sum_{u''\in \mathcal{U}} d_{u''} \right)$ 
triplets for user $u$, where $S_u$ is a weighted fraction of the total number of interactions, rounded to the nearest integer.
The hyperparameter $\alpha$ tunes the significance of high-activity users. When $\alpha=0$, all users have the same number of training samples per epoch, regardless of the number of observed positives. This approach is used in LightGCN~\cite{he2020lightgcn}. When $\alpha=1$, the number of samples for user $u$ is equal to the number of observed interactions, $d_u$. This approach is most faithful to a sample-based approximation of the BPR loss function, Eq.~\eqref{eqn:bpr}.\footnote{See Appendix \ref{sec:BPR-sampling} for a more in-depth discussion.} The normalizing denominator and the right term denoting the total number of interactions in $S_u$ are constants and are not included in the loss function in Eq. \eqref{eqn:our-loss}.

\subsection{Relationship to Prior Work} \label{sec:prior-work}

Our approach to reweighting items is related to causal approaches to accounting for popularity bias \cite{schnabel2016recommendations}. These works recognize that the unobserved interactions in $\mathbf{Y}$ are not missing at random. A user interacts with an item only when exposed, so a missing interaction with an item does not necessarily imply a lack of relevance. A common debiasing approach is Inverse Propensity Weighting (IPW). In IPW, for a user $u$, an item $i$ has a latent propensity probability $w_{ui} \in [0, 1]$ of being exposed to the user, so to debias the estimator, the loss for $u, i$ is reweighted by $1/w_{ui}$. Although there has been previous work on training a separate model to estimate propensity, there are also heuristic-based approaches. Our item reweighting approach is equivalent to prior popularity-heuristic approaches to Inverse Propensity Weighting \cite{lee2021dual, saito2020unbiased} where the propensity for an item $i$ is proportional to $d_i^{-\beta}$, where a common setting is $\beta=1/2$. 

We also note that past work has proposed alternative modifications to the BPR loss to mitigate popularity bias \cite{rhee2022countering}. 
In contrast to existing reweighting approaches, \texttt{PAIR}'s novelty lies in the empirical identification and valuation of power-niche users, which directly motivates the $\alpha,\beta$ formulation.

%% file: sections/6_experiments.tex
\section{Experiments} \label{sec:eval}
We present an empirical evaluation of our reweighting framework \texttt{PAIR}, instantiated with LightGCN and matrix factorization as base models.
Our results show that by simultaneously accounting for user activity level \textit{and} item-popularity heterogeneities, we are able to Pareto-dominate vanilla models, increasing recommendation performance and decreasing popularity bias. We also show that \texttt{PAIR} yields superior performance compared to approaches that consider only item popularity.

\subsection{Methodology} \label{sec:methodology}

\paragraph{Base Model}
We define a \emph{base model} as any score prediction model that produces a score prediction matrix $\mathbf{\hat{Y}}\in\{0, 1\}^{n\times m}$. Our reweighting framework is specific to the BPR loss function and can be applied to any base model. In the following results, we utilize two base models: Matrix Factorization (MF) and LightGCN (LGN) \cite{he2020lightgcn}, a deep collaborative filtering model. We instantiate LightGCN with $3$ layers and use the hyperparameters reported in \citet{he2020lightgcn}. The MF base model defines the prediction score for user $u$ and item $i$ as
$\mathbf{\hat{Y}}_{ui} = \sigma\left(\mathbf{u}_u^T\mathbf{v}_i\right)$, 
where $\mathbf{u}_u, \mathbf{v}_i \in \mathbb{R}^{d}$ are $d$-dimensional user and item embeddings. Both base models also introduce a regularization on the Frobenius norm of the embeddings.  

\paragraph{Vanilla Models} Past sampling implementations of the BPR loss have implicitly reweighted users based on activity level, while \texttt{PAIR} explicitly tunes the activity-level reweighting. The two common implementations of the BPR loss are specific instances of the $\alpha,\beta$ framework in \texttt{PAIR}. The first instance is the most closely aligned with sample-based approximations of the BPR loss and sets $\alpha=1, \beta=0$, where the number of positive samples for a user is proportional to the user's activity level. The second instance, as in the case of LightGCN, sets $\alpha=0, \beta=0$ such that all users have the same number of positive samples. Sampling a fixed number of positives for all users implicitly upweights light users. In our evaluation, we compare against both implementations of the BPR loss where ``Vanilla'' refers to $\alpha=0$, to reproduce the results in \citet{he2020lightgcn}, and ``Vanilla-BPR'' refers to $\alpha=1$, to indicate alignment with the original BPR loss in Eq. \eqref{eqn:bpr}.

\paragraph{Baselines}
We compare against representative baselines from three popularity-bias mitigation categories:
\begin{itemize}[leftmargin=*, topsep=2pt, itemsep=1pt, parsep=0pt]
    \item \textbf{Regularization (``Pop. Reg.'')} We compare against the popularity-regularization mitigation method introduced in \citet{zhu2021popularityopportunity}. The regularization penalizes positive correlations between prediction scores and item popularity. There is a hyperparameter $\gamma > 0$ to control the regularization strength. 
    \item \textbf{Re-weighting (IPW)} We compare against the commonly-used Inverse Propensity Weighting approach for mitigating popularity bias as introduced in \citet{schnabel2016recommendations}. To account for unequal item exposure probabilities, IPW reweights an interaction between user $u$ and item $j$ by $p_j^{-\beta}$ where $p_j$ is the \emph{propensity} of item $j$. Similar to past heuristic approaches to estimating propensity, an item's propensity is equal to its popularity in our experiments, and we tune the propensity exponent hyperparameter $\beta$. This formulation of IPW is equivalent to fixing $\alpha=0$ in \texttt{PAIR}.
    \item \textbf{Post-process Re-ranking (``Pop. Comp.'')} We compare against the popularity-compensation post-processing technique introduced in \citet{zhu2021popularityopportunity}. The technique interpolates the raw prediction scores $\mathbf{\hat{Y}}$ with a popularity compensation, where less popular items receive greater compensation. Pop. Comp. utilizes two hyperparameters, not to be confused with those in \texttt{PAIR}, $\alpha', \beta'$, where $\alpha'$ dictates the linear interpolation and $\beta'$ dictates the amount of compensation.
\end{itemize}

\paragraph{Data and Metrics} We run our experiments on the three datasets introduced in Sec. \ref{sec:data-analysis}, using the train/validation/test splits introduced in \citet{lee2024revisiting}. We evaluate recommendation performance via the standard Recall@$k$, Precision@$k$, and NDCG@$k$ ranking metrics, where $k=20$. We measure popularity bias with both item and user-centric metrics. For item-centric metrics, we use the Popularity-Opportunity Bias (POB) metric~\cite{zhu2021popularityopportunity}, which is the correlation between an item's popularity and its average ranking among users who interact with the item, and Coverage, which denotes the fraction of all items that are ranked in the top $k$ at least once among all users. For a user-centric perspective, we also measure the Recall@$k$ for niche users, which we refer to as the Niche Recall@$k$. 

\paragraph{Hyperparameter Optimization} We select the $\alpha, \beta$ pair that maximizes \texttt{PAIR}'s Recall@$k$ on the validation set. Optimized values and additional hyperparameter optimization methodology for \texttt{PAIR} and the baselines are included in Appendix \ref{sec:optimized-hyperparam}.\footnote{Our code is available at \href{https://github.com/dliu18/power-niche-users-www}{https://github.com/dliu18/power-niche-users-www}.}

\subsection{Results} \label{sec:results}

\input{sections/tables-and-figures/performance_vs_vanilla_with_CIs}

\textbf{Result 1: \texttt{PAIR} upweights power-niche users the most relative to BPR loss.} 
Fig. \ref{fig:quadrant-weight-changes} confirms that using the hyperparameter-optimized $\alpha$ and $\beta$, \texttt{PAIR} upweights power-niche users. The weight of a quadrant is the number of positive samples drawn for all users in that quadrant. Under Vanilla, with $\alpha=0$, a fixed number of positive samples is drawn per user, so the weight of a quadrant is proportional to the number of users in that quadrant. \texttt{PAIR} weighs user-item samples by item popularity, where a sample $u, i$ is equivalent to $d_i^{-\beta}$ unweighted samples. Thus, the expected weight for user $u$ is proportional to ${d_u}^\alpha\left(\sum_{i\in N_u}d_i^{-\beta} / d_u\right)$. The values in Fig. \ref{fig:quadrant-weight-changes} indicate the percentage weight change for each quadrant, using the expected weights for \texttt{PAIR}. In all dataset-model configurations, power-niche users are upweighted.

\input{sections/tables-and-figures/quadrant-weight-changes}

\input{sections/tables-and-figures/ipw-baseline}

\textbf{Result 2: Increasing the weight of power-niche users increases recommendation performance while decreasing popularity bias.} 
Table \ref{tab:results} lists the evaluation metrics for all model variants across all dataset and model configurations. The reported means and confidence intervals are calculated over four trials.
The results show that tuning $\alpha,\beta$ based on validation Recall@$k$ results in not only improved performance but also mitigated bias on the test set. 
\texttt{PAIR} Pareto-dominates the Vanilla model in five of the six configurations, yielding higher performance across the downstream metrics and mitigating popularity bias according to the bias metrics. In the one instance where \texttt{PAIR} does not Pareto-dominate, LightGCN trained on Gowalla, our Recall@$k$ is $0.0008$ less than the baseline.
On average, for matrix factorization, \texttt{PAIR} increases Recall@$k$ by $20.1\%$, Precision@$k$ by $17.7\%$, and NDCG by $16.9\%$ while reducing POB by $7.5\%$, increasing coverage by $9.4\%$, and increasing Niche Recall@$k$ by $23.0\%$. The trend persists with LightGCN but with a more muted impact. For instance, in LightGCN, \texttt{PAIR} increases Recall@$k$ by $1.2\%$ while mitigating popularity bias. 

Comparing with baselines, IPW Pareto-dominates Vanilla in all six configurations; however, IPW's downstream performance is not as strong, with \texttt{PAIR} achieving the highest downstream performance in three of the six configurations.
In the remaining three configurations, the top-performing model is still an instance of \texttt{PAIR}'s $\alpha,\beta$ framework (Vanilla models and IPW), indicating that the $\alpha,\beta$ pairs that maximize validation Recall@$k$ in these configurations do not generalize to top test-set performance.

Pop. Comp. is effective in reducing POB, and Pop. Reg. is particularly effective for increasing coverage, but both of these baselines mitigate popularity bias at a high cost to performance. For instance, on average, Popularity Compensation reduces Recall@$k$ by $10.2\%$.

\textbf{Result 3: Considering only item propensity sacrifices downstream performance.}
Given IPW's strong performance among the three baselines, we provide a more in-depth comparison between \texttt{PAIR} and IPW, where the methodological difference is \texttt{PAIR}'s consideration of user activity level in addition to item propensity. Fig. \ref{fig:ipw-baseline} shows that across hyperparameter configurations, IPW suffers from a performance-bias trade-off for LightGCN. We train IPW baseline models for increasing values of $\beta$, where increasing $\beta$ upweights less popular items. As expected, popularity bias reduces as $\beta$ increases. However, for LightGCN, increasing $\beta$ initially preserves performance before degrading, illustrating the bias-performance trade-off for IPW. In contrast, for Yelp2018 and Amazon-Book, \texttt{PAIR} noticeably improves performance. The results in Fig. \ref{fig:ipw-baseline} reflect the performance of a single run of each model.

%% file: sections/tables-and-figures/performance_vs_vanilla_with_CIs.tex
\begin{table*}[ht]
\centering
\caption{Evaluation metrics for \texttt{PAIR} compared to the vanilla and baseline bias-mitigation methods.
Highlighted rows denote the model variant with the highest Recall for a given dataset and model combination. In five of six dataset-model configurations, \texttt{PAIR} Pareto-dominates Vanilla with higher Recall and lower Popularity-Opportunity Bias (POB).}
\footnotesize
\begin{tabular}{lllcccccc}
\toprule
\textbf{Dataset} & \textbf{Model} & \textbf{Variant} & \textbf{Recall@20 ($\uparrow$)} & \textbf{Precision@20 ($\uparrow$)} & \textbf{NDCG@20 ($\uparrow$)} & \textbf{POB@20 ($\downarrow$)} & \textbf{Coverage@20 ($\uparrow$)} & \textbf{Niche Recall@20 ($\uparrow$)} \\
\midrule
Amazon-Book & MF & Vanilla & 0.0039{\scriptsize\,(\,$\pm$\,0.0002\,)} & 0.0025{\scriptsize\,(\,$\pm$\,0.0001\,)} & 0.0035{\scriptsize\,(\,$\pm$\,0.0002\,)} & 0.4862{\scriptsize\,(\,$\pm$\,0.0063\,)} & 46.2\%{\scriptsize\,(\,$\pm$\,1.4\%\,)} & 0.0022{\scriptsize\,(\,$\pm$\,0.0003\,)} \\
 \bestrow &  & Vanilla-BPR & \textbf{\underline{0.0044{\scriptsize\,(\,$\pm$\,0.0002\,)}}} & \textbf{\underline{0.0030{\scriptsize\,(\,$\pm$\,0.0002\,)}}} & \textbf{\underline{0.0041{\scriptsize\,(\,$\pm$\,0.0002\,)}}} & 0.4774{\scriptsize\,(\,$\pm$\,0.0050\,)} & 54.0\%{\scriptsize\,(\,$\pm$\,0.8\%\,)} & \textbf{\underline{0.0029{\scriptsize\,(\,$\pm$\,0.0003\,)}}} \\
 &  & \texttt{PAIR} & 0.0043{\scriptsize\,(\,$\pm$\,0.0002\,)} & 0.0030{\scriptsize\,(\,$\pm$\,0.0002\,)} & 0.0040{\scriptsize\,(\,$\pm$\,0.0002\,)} & 0.4844{\scriptsize\,(\,$\pm$\,0.0067\,)} & 51.8\%{\scriptsize\,(\,$\pm$\,0.8\%\,)} & 0.0028{\scriptsize\,(\,$\pm$\,0.0004\,)} \\
 &  & IPW & 0.0040{\scriptsize\,(\,$\pm$\,0.0002\,)} & 0.0026{\scriptsize\,(\,$\pm$\,0.0001\,)} & 0.0036{\scriptsize\,(\,$\pm$\,0.0002\,)} & 0.4771{\scriptsize\,(\,$\pm$\,0.0059\,)} & 49.6\%{\scriptsize\,(\,$\pm$\,1.3\%\,)} & 0.0022{\scriptsize\,(\,$\pm$\,0.0003\,)} \\
 &  & Pop. Comp. & 0.0020{\scriptsize\,(\,$\pm$\,0.0001\,)} & 0.0018{\scriptsize\,(\,$\pm$\,0.0001\,)} & 0.0021{\scriptsize\,(\,$\pm$\,0.0001\,)} & 0.4865{\scriptsize\,(\,$\pm$\,0.0061\,)} & \textbf{\underline{92.4\%{\scriptsize\,(\,$\pm$\,0.1\%\,)}}} & 0.0015{\scriptsize\,(\,$\pm$\,0.0002\,)} \\
 &  & Pop. Reg. & 0.0017{\scriptsize\,(\,$\pm$\,0.0001\,)} & 0.0010{\scriptsize\,(\,$\pm$\,0.0000\,)} & 0.0015{\scriptsize\,(\,$\pm$\,0.0000\,)} & \textbf{\underline{0.4250{\scriptsize\,(\,$\pm$\,0.0086\,)}}} & 70.2\%{\scriptsize\,(\,$\pm$\,0.4\%\,)} & 0.0012{\scriptsize\,(\,$\pm$\,0.0001\,)} \\
\cmidrule(lr){3-9}
Gowalla & MF & Vanilla & 0.0976{\scriptsize\,(\,$\pm$\,0.0009\,)} & 0.0293{\scriptsize\,(\,$\pm$\,0.0001\,)} & 0.0747{\scriptsize\,(\,$\pm$\,0.0005\,)} & 0.1475{\scriptsize\,(\,$\pm$\,0.0065\,)} & 66.0\%{\scriptsize\,(\,$\pm$\,0.2\%\,)} & 0.1201{\scriptsize\,(\,$\pm$\,0.0023\,)} \\
 &  & Vanilla-BPR & 0.0910{\scriptsize\,(\,$\pm$\,0.0015\,)} & 0.0278{\scriptsize\,(\,$\pm$\,0.0003\,)} & 0.0693{\scriptsize\,(\,$\pm$\,0.0014\,)} & 0.1622{\scriptsize\,(\,$\pm$\,0.0067\,)} & 69.6\%{\scriptsize\,(\,$\pm$\,0.1\%\,)} & 0.1095{\scriptsize\,(\,$\pm$\,0.0033\,)} \\
\bestrow &  & \texttt{PAIR} & \textbf{\underline{0.1203{\scriptsize\,(\,$\pm$\,0.0011\,)}}} & \textbf{\underline{0.0350{\scriptsize\,(\,$\pm$\,0.0002\,)}}} & \textbf{\underline{0.0867{\scriptsize\,(\,$\pm$\,0.0010\,)}}} & \textbf{\underline{0.1243{\scriptsize\,(\,$\pm$\,0.0054\,)}}} & 69.7\%{\scriptsize\,(\,$\pm$\,0.2\%\,)} & \textbf{\underline{0.1401{\scriptsize\,(\,$\pm$\,0.0019\,)}}} \\
 &  & IPW & 0.1203{\scriptsize\,(\,$\pm$\,0.0011\,)} & 0.0350{\scriptsize\,(\,$\pm$\,0.0002\,)} & 0.0867{\scriptsize\,(\,$\pm$\,0.0010\,)} & 0.1243{\scriptsize\,(\,$\pm$\,0.0054\,)} & 69.7\%{\scriptsize\,(\,$\pm$\,0.2\%\,)} & 0.1401{\scriptsize\,(\,$\pm$\,0.0019\,)} \\
 &  & Pop. Comp. & 0.0965{\scriptsize\,(\,$\pm$\,0.0012\,)} & 0.0286{\scriptsize\,(\,$\pm$\,0.0002\,)} & 0.0715{\scriptsize\,(\,$\pm$\,0.0007\,)} & 0.1472{\scriptsize\,(\,$\pm$\,0.0062\,)} & 67.9\%{\scriptsize\,(\,$\pm$\,0.3\%\,)} & 0.1197{\scriptsize\,(\,$\pm$\,0.0031\,)} \\
 &  & Pop. Reg. & 0.0316{\scriptsize\,(\,$\pm$\,0.0011\,)} & 0.0094{\scriptsize\,(\,$\pm$\,0.0003\,)} & 0.0206{\scriptsize\,(\,$\pm$\,0.0008\,)} & 0.2913{\scriptsize\,(\,$\pm$\,0.0127\,)} & \textbf{\underline{79.3\%{\scriptsize\,(\,$\pm$\,0.5\%\,)}}} & 0.0377{\scriptsize\,(\,$\pm$\,0.0014\,)} \\
\cmidrule(lr){3-9}
Yelp2018 & MF & Vanilla & 0.0276{\scriptsize\,(\,$\pm$\,0.0015\,)} & 0.0133{\scriptsize\,(\,$\pm$\,0.0006\,)} & 0.0224{\scriptsize\,(\,$\pm$\,0.0009\,)} & 0.3717{\scriptsize\,(\,$\pm$\,0.0042\,)} & 63.2\%{\scriptsize\,(\,$\pm$\,0.2\%\,)} & 0.0273{\scriptsize\,(\,$\pm$\,0.0014\,)} \\
 &  & Vanilla-BPR & 0.0277{\scriptsize\,(\,$\pm$\,0.0008\,)} & 0.0134{\scriptsize\,(\,$\pm$\,0.0003\,)} & 0.0222{\scriptsize\,(\,$\pm$\,0.0006\,)} & 0.4060{\scriptsize\,(\,$\pm$\,0.0014\,)} & 67.1\%{\scriptsize\,(\,$\pm$\,0.2\%\,)} & 0.0266{\scriptsize\,(\,$\pm$\,0.0007\,)} \\
 &  & \texttt{PAIR} & 0.0344{\scriptsize\,(\,$\pm$\,0.0007\,)} & 0.0155{\scriptsize\,(\,$\pm$\,0.0003\,)} & 0.0267{\scriptsize\,(\,$\pm$\,0.0006\,)} & 0.3483{\scriptsize\,(\,$\pm$\,0.0046\,)} & 69.8\%{\scriptsize\,(\,$\pm$\,0.1\%\,)} & 0.0347{\scriptsize\,(\,$\pm$\,0.0010\,)} \\
\bestrow &  & IPW & \textbf{\underline{0.0350{\scriptsize\,(\,$\pm$\,0.0004\,)}}} & \textbf{\underline{0.0157{\scriptsize\,(\,$\pm$\,0.0001\,)}}} & \textbf{\underline{0.0272{\scriptsize\,(\,$\pm$\,0.0005\,)}}} & \textbf{\underline{0.3429{\scriptsize\,(\,$\pm$\,0.0015\,)}}} & 66.8\%{\scriptsize\,(\,$\pm$\,0.2\%\,)} & \textbf{\underline{0.0356{\scriptsize\,(\,$\pm$\,0.0008\,)}}} \\
 &  & Pop. Comp. & 0.0246{\scriptsize\,(\,$\pm$\,0.0011\,)} & 0.0118{\scriptsize\,(\,$\pm$\,0.0004\,)} & 0.0188{\scriptsize\,(\,$\pm$\,0.0007\,)} & 0.3707{\scriptsize\,(\,$\pm$\,0.0043\,)} & \textbf{\underline{79.8\%{\scriptsize\,(\,$\pm$\,0.5\%\,)}}} & 0.0248{\scriptsize\,(\,$\pm$\,0.0009\,)} \\
 &  & Pop. Reg. & 0.0120{\scriptsize\,(\,$\pm$\,0.0004\,)} & 0.0053{\scriptsize\,(\,$\pm$\,0.0001\,)} & 0.0088{\scriptsize\,(\,$\pm$\,0.0002\,)} & 0.4061{\scriptsize\,(\,$\pm$\,0.0066\,)} & 75.6\%{\scriptsize\,(\,$\pm$\,0.8\%\,)} & 0.0140{\scriptsize\,(\,$\pm$\,0.0007\,)} \\
\cmidrule(lr){3-9}
Amazon-Book & LGN & Vanilla & 0.0349{\scriptsize\,(\,$\pm$\,0.0002\,)} & 0.0144{\scriptsize\,(\,$\pm$\,0.0000\,)} & 0.0269{\scriptsize\,(\,$\pm$\,0.0001\,)} & 0.4169{\scriptsize\,(\,$\pm$\,0.0021\,)} & 24.3\%{\scriptsize\,(\,$\pm$\,0.2\%\,)} & 0.0315{\scriptsize\,(\,$\pm$\,0.0003\,)} \\
 &  & Vanilla-BPR & 0.0351{\scriptsize\,(\,$\pm$\,0.0001\,)} & \textbf{\underline{0.0149{\scriptsize\,(\,$\pm$\,0.0000\,)}}} & \textbf{\underline{0.0273{\scriptsize\,(\,$\pm$\,0.0001\,)}}} & 0.4062{\scriptsize\,(\,$\pm$\,0.0011\,)} & 24.4\%{\scriptsize\,(\,$\pm$\,0.3\%\,)} & \textbf{\underline{0.0318{\scriptsize\,(\,$\pm$\,0.0002\,)}}} \\
\bestrow &  & \texttt{PAIR} & \textbf{\underline{0.0352{\scriptsize\,(\,$\pm$\,0.0002\,)}}} & 0.0148{\scriptsize\,(\,$\pm$\,0.0001\,)} & 0.0273{\scriptsize\,(\,$\pm$\,0.0002\,)} & \textbf{\underline{0.3783{\scriptsize\,(\,$\pm$\,0.0016\,)}}} & 25.5\%{\scriptsize\,(\,$\pm$\,0.2\%\,)} & 0.0316{\scriptsize\,(\,$\pm$\,0.0001\,)} \\
 &  & IPW & 0.0351{\scriptsize\,(\,$\pm$\,0.0002\,)} & 0.0146{\scriptsize\,(\,$\pm$\,0.0001\,)} & 0.0271{\scriptsize\,(\,$\pm$\,0.0001\,)} & 0.4039{\scriptsize\,(\,$\pm$\,0.0021\,)} & 24.8\%{\scriptsize\,(\,$\pm$\,0.2\%\,)} & 0.0316{\scriptsize\,(\,$\pm$\,0.0004\,)} \\
 &  & Pop. Comp. & 0.0351{\scriptsize\,(\,$\pm$\,0.0003\,)} & 0.0147{\scriptsize\,(\,$\pm$\,0.0001\,)} & 0.0272{\scriptsize\,(\,$\pm$\,0.0002\,)} & 0.4056{\scriptsize\,(\,$\pm$\,0.0023\,)} & \textbf{\underline{30.2\%{\scriptsize\,(\,$\pm$\,0.2\%\,)}}} & 0.0317{\scriptsize\,(\,$\pm$\,0.0006\,)} \\
 &  & Pop. Reg. & 0.0344{\scriptsize\,(\,$\pm$\,0.0004\,)} & 0.0141{\scriptsize\,(\,$\pm$\,0.0002\,)} & 0.0265{\scriptsize\,(\,$\pm$\,0.0003\,)} & 0.4168{\scriptsize\,(\,$\pm$\,0.0014\,)} & 25.6\%{\scriptsize\,(\,$\pm$\,0.3\%\,)} & 0.0312{\scriptsize\,(\,$\pm$\,0.0005\,)} \\
\cmidrule(lr){3-9}
\bestrow Gowalla & LGN & Vanilla & \textbf{\underline{0.1805{\scriptsize\,(\,$\pm$\,0.0007\,)}}} & 0.0551{\scriptsize\,(\,$\pm$\,0.0001\,)} & 0.1532{\scriptsize\,(\,$\pm$\,0.0007\,)} & 0.1635{\scriptsize\,(\,$\pm$\,0.0034\,)} & 41.5\%{\scriptsize\,(\,$\pm$\,0.2\%\,)} & \textbf{\underline{0.2006{\scriptsize\,(\,$\pm$\,0.0015\,)}}} \\
 &  & Vanilla-BPR & 0.1766{\scriptsize\,(\,$\pm$\,0.0010\,)} & 0.0544{\scriptsize\,(\,$\pm$\,0.0002\,)} & 0.1507{\scriptsize\,(\,$\pm$\,0.0004\,)} & 0.1642{\scriptsize\,(\,$\pm$\,0.0030\,)} & \textbf{\underline{50.3\%{\scriptsize\,(\,$\pm$\,0.3\%\,)}}} & 0.1974{\scriptsize\,(\,$\pm$\,0.0005\,)} \\
 &  & \texttt{PAIR} & 0.1797{\scriptsize\,(\,$\pm$\,0.0007\,)} & \textbf{\underline{0.0553{\scriptsize\,(\,$\pm$\,0.0002\,)}}} & \textbf{\underline{0.1536{\scriptsize\,(\,$\pm$\,0.0005\,)}}} & \textbf{\underline{0.1605{\scriptsize\,(\,$\pm$\,0.0019\,)}}} & 45.8\%{\scriptsize\,(\,$\pm$\,0.3\%\,)} & 0.2005{\scriptsize\,(\,$\pm$\,0.0012\,)} \\
 &  & IPW & 0.1804{\scriptsize\,(\,$\pm$\,0.0006\,)} & 0.0552{\scriptsize\,(\,$\pm$\,0.0001\,)} & 0.1530{\scriptsize\,(\,$\pm$\,0.0005\,)} & 0.1615{\scriptsize\,(\,$\pm$\,0.0030\,)} & 38.8\%{\scriptsize\,(\,$\pm$\,0.2\%\,)} & 0.2001{\scriptsize\,(\,$\pm$\,0.0010\,)} \\
 &  & Pop. Comp. & 0.1798{\scriptsize\,(\,$\pm$\,0.0007\,)} & 0.0546{\scriptsize\,(\,$\pm$\,0.0001\,)} & 0.1524{\scriptsize\,(\,$\pm$\,0.0006\,)} & 0.1618{\scriptsize\,(\,$\pm$\,0.0032\,)} & 44.1\%{\scriptsize\,(\,$\pm$\,0.2\%\,)} & 0.1997{\scriptsize\,(\,$\pm$\,0.0015\,)} \\
 &  & Pop. Reg. & 0.1760{\scriptsize\,(\,$\pm$\,0.0008\,)} & 0.0539{\scriptsize\,(\,$\pm$\,0.0001\,)} & 0.1481{\scriptsize\,(\,$\pm$\,0.0006\,)} & 0.1637{\scriptsize\,(\,$\pm$\,0.0022\,)} & 43.6\%{\scriptsize\,(\,$\pm$\,0.1\%\,)} & 0.1951{\scriptsize\,(\,$\pm$\,0.0008\,)} \\
\cmidrule(lr){3-9}
Yelp2018 & LGN & Vanilla & 0.0620{\scriptsize\,(\,$\pm$\,0.0004\,)} & 0.0278{\scriptsize\,(\,$\pm$\,0.0002\,)} & 0.0507{\scriptsize\,(\,$\pm$\,0.0003\,)} & 0.4079{\scriptsize\,(\,$\pm$\,0.0021\,)} & 39.8\%{\scriptsize\,(\,$\pm$\,0.2\%\,)} & 0.0628{\scriptsize\,(\,$\pm$\,0.0006\,)} \\
 &  & Vanilla-BPR & 0.0629{\scriptsize\,(\,$\pm$\,0.0006\,)} & 0.0283{\scriptsize\,(\,$\pm$\,0.0002\,)} & 0.0515{\scriptsize\,(\,$\pm$\,0.0002\,)} & 0.4043{\scriptsize\,(\,$\pm$\,0.0019\,)} & \textbf{\underline{43.4\%{\scriptsize\,(\,$\pm$\,0.4\%\,)}}} & 0.0641{\scriptsize\,(\,$\pm$\,0.0009\,)} \\
\bestrow &  & \texttt{PAIR} & \textbf{\underline{0.0641{\scriptsize\,(\,$\pm$\,0.0001\,)}}} & \textbf{\underline{0.0288{\scriptsize\,(\,$\pm$\,0.0001\,)}}} & \textbf{\underline{0.0522{\scriptsize\,(\,$\pm$\,0.0002\,)}}} & \textbf{\underline{0.3805{\scriptsize\,(\,$\pm$\,0.0017\,)}}} & 40.9\%{\scriptsize\,(\,$\pm$\,0.5\%\,)} & \textbf{\underline{0.0655{\scriptsize\,(\,$\pm$\,0.0006\,)}}} \\
 &  & IPW & 0.0630{\scriptsize\,(\,$\pm$\,0.0003\,)} & 0.0282{\scriptsize\,(\,$\pm$\,0.0002\,)} & 0.0510{\scriptsize\,(\,$\pm$\,0.0002\,)} & 0.3966{\scriptsize\,(\,$\pm$\,0.0011\,)} & 35.2\%{\scriptsize\,(\,$\pm$\,0.3\%\,)} & 0.0640{\scriptsize\,(\,$\pm$\,0.0004\,)} \\
 &  & Pop. Comp. & 0.0621{\scriptsize\,(\,$\pm$\,0.0005\,)} & 0.0278{\scriptsize\,(\,$\pm$\,0.0001\,)} & 0.0507{\scriptsize\,(\,$\pm$\,0.0003\,)} & 0.4059{\scriptsize\,(\,$\pm$\,0.0029\,)} & 41.0\%{\scriptsize\,(\,$\pm$\,0.1\%\,)} & 0.0628{\scriptsize\,(\,$\pm$\,0.0008\,)} \\
 &  & Pop. Reg. & 0.0603{\scriptsize\,(\,$\pm$\,0.0005\,)} & 0.0270{\scriptsize\,(\,$\pm$\,0.0001\,)} & 0.0488{\scriptsize\,(\,$\pm$\,0.0004\,)} & 0.4072{\scriptsize\,(\,$\pm$\,0.0018\,)} & 40.8\%{\scriptsize\,(\,$\pm$\,0.3\%\,)} & 0.0615{\scriptsize\,(\,$\pm$\,0.0006\,)} \\
\cmidrule(lr){3-9}
\bottomrule
\end{tabular}%

\label{tab:results}
\end{table*}

%% file: sections/tables-and-figures/quadrant-weight-changes.tex
\begin{figure}
    \centering
    \begin{subfigure}{\columnwidth}
        \centering
        \includegraphics[width=\columnwidth]{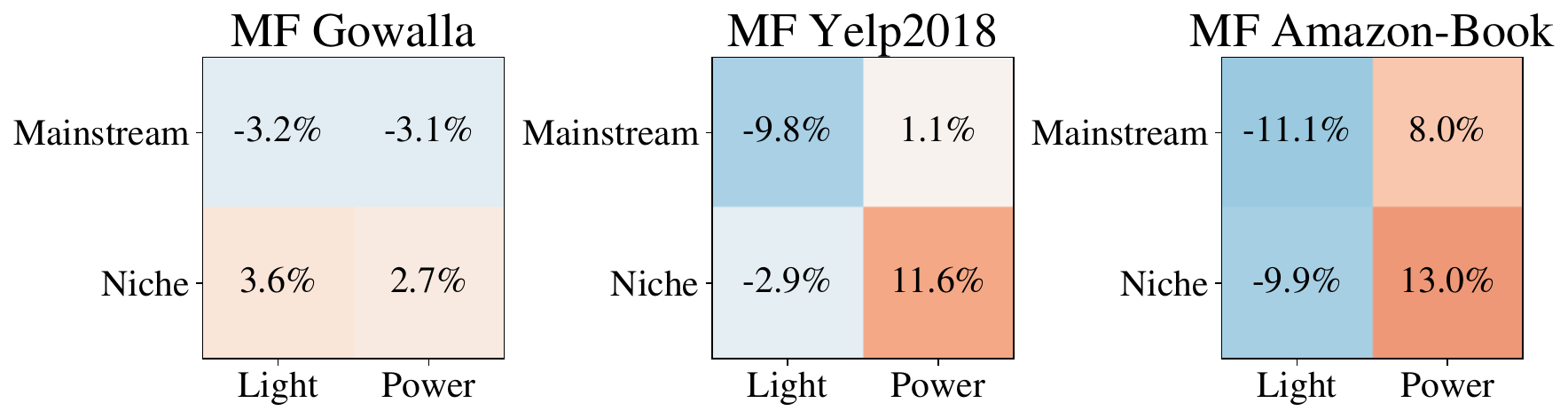}
    \end{subfigure}
    \begin{subfigure}{\columnwidth}
        \centering
        \includegraphics[width=\columnwidth]{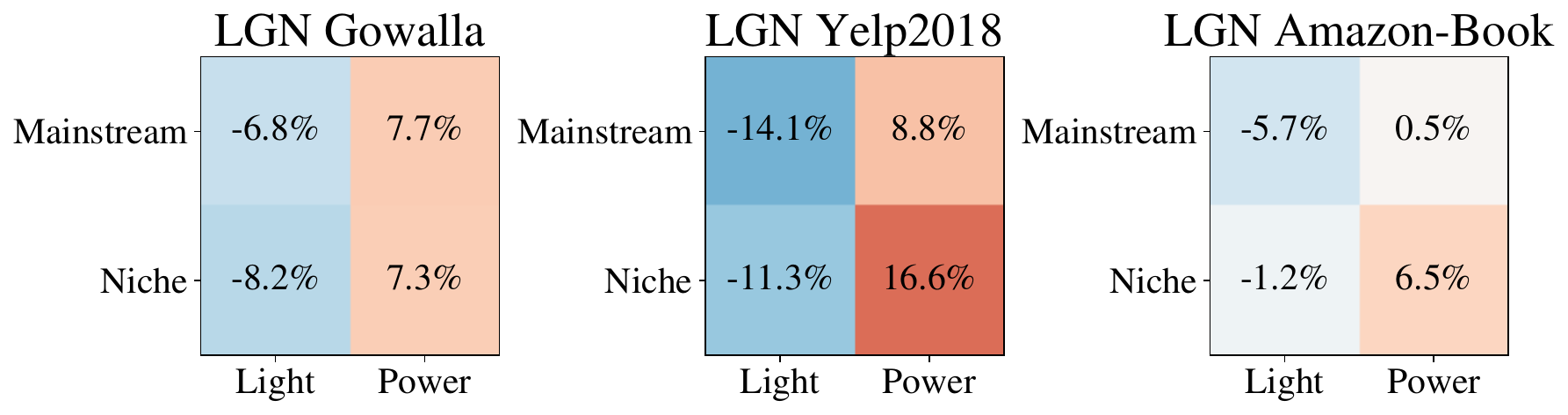}
    \end{subfigure}
    \Description{Heatmaps showing the percentage change in total training weight assigned to each user quadrant under PAIR compared to the baseline BPR loss. Results are shown for matrix factorization and LightGCN across datasets, indicating that power-niche users receive the largest increase in weight while light users are downweighted.}
    \caption{Under \texttt{PAIR}, our reweighting framework, power-niche users are upweighted the most relative to the BPR loss. The weight of user $u$ is the number of positive items sampled per epoch.
    The percentages in the heatmaps indicate the change in the relative total weight of each quadrant when applying the optimized hyperparameters for \texttt{PAIR} to MF and LGN; hence, the percentages sum to zero in each heatmap. In contrast to power-niche users, light users exhibit a decrease in weight, and power-mainstream users exhibit a marginal increase in weight.}
    \label{fig:quadrant-weight-changes}
\end{figure}

%% file: sections/tables-and-figures/ipw-baseline.tex
\begin{figure}
    \centering
    \begin{subfigure}{\columnwidth}
        \centering
        \includegraphics[width=\columnwidth]{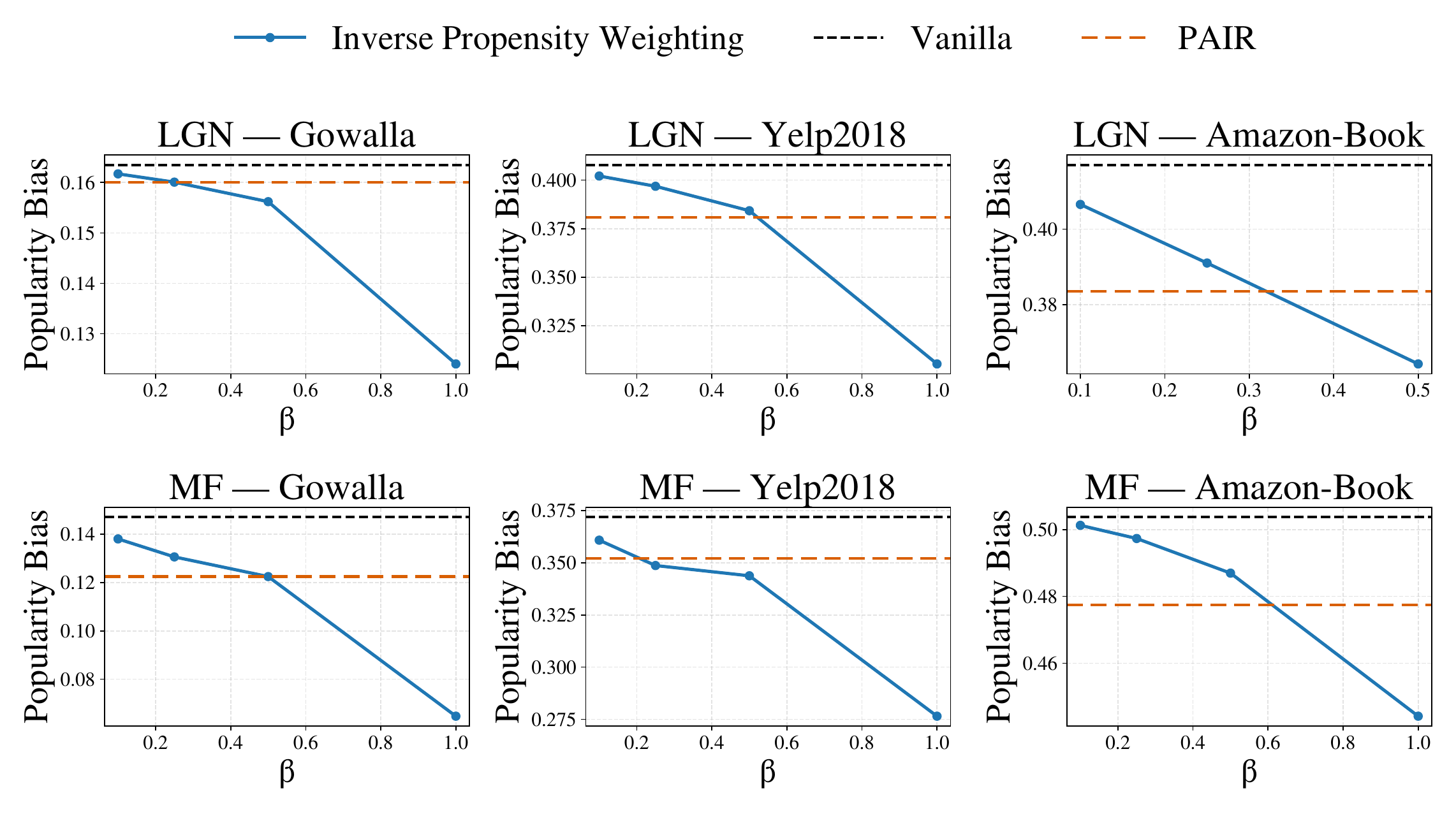}
    \end{subfigure}
    \begin{subfigure}{\columnwidth}
        \centering
        \includegraphics[width=\columnwidth]{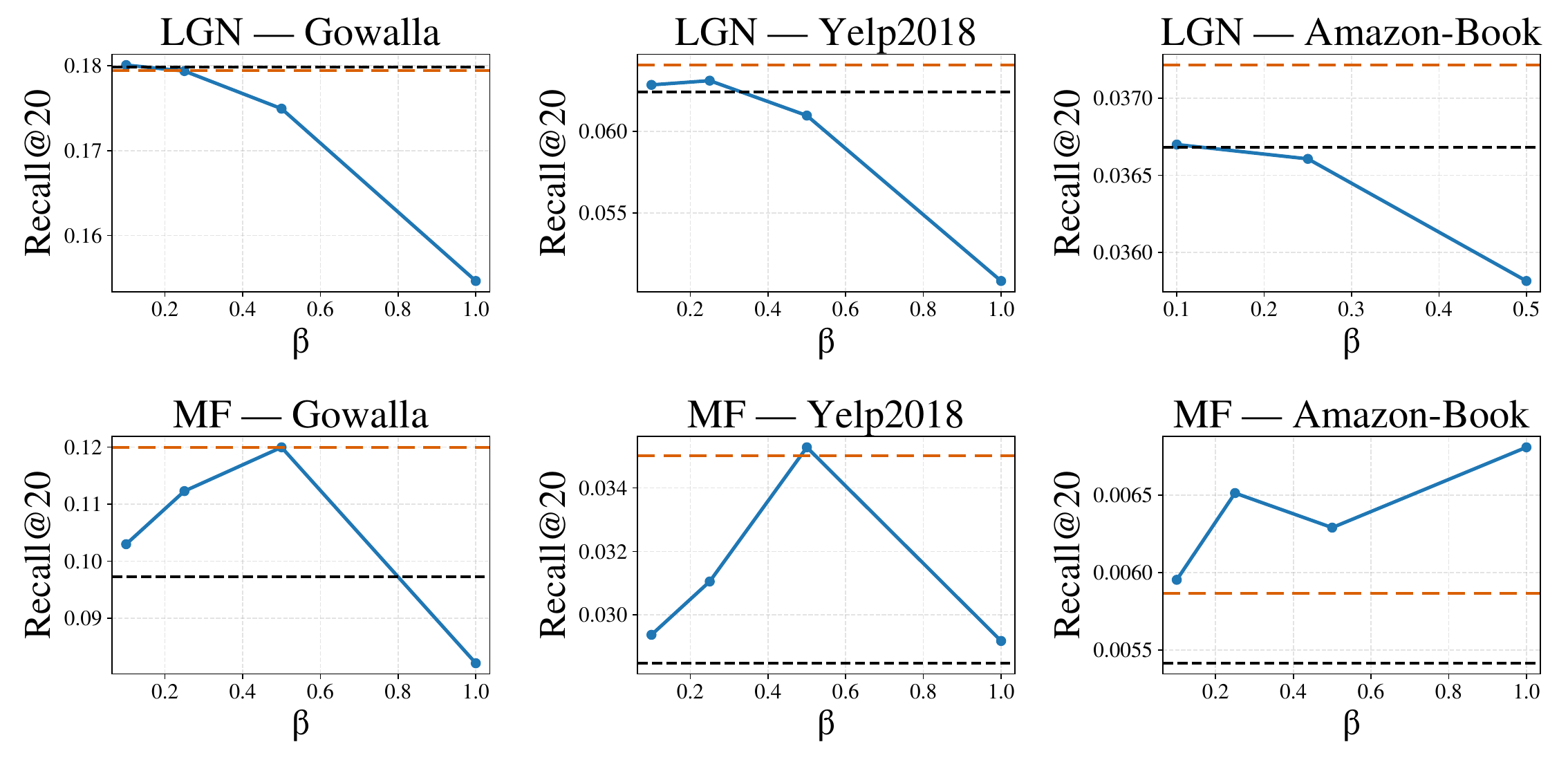}
    \end{subfigure}
    \Description{Plots comparing PAIR and inverse propensity weighting across a range of item-popularity reweighting parameters. The top rows show changes in popularity bias, and the bottom rows show Recall@20 for matrix factorization and LightGCN models across datasets. PAIR consistently improves performance while reducing bias, whereas IPW exhibits a bias–performance trade-off for deep models.}
    \caption{We provide a more in-depth comparison between \texttt{PAIR} and Inverse Propensity Weighting (IPW), which reweights based on item popularity but not user activity level.
    The above plots illustrate IPW's performance over a spectrum of $\beta$. As $\beta$ increases, less popular items are upweighted more. The top two rows indicate popularity bias, as measured with POB, while the bottom two rows indicate Recall@$k$.
    The plots show that, for shallow matrix factorization (MF), IPW Pareto-dominates Vanilla (decreasing popularity bias while improving performance), whereas for the deep LightGCN model, IPW presents a trade-off. In contrast, \texttt{PAIR} considers user activity level in addition to item propensity and consistently Pareto-dominates the vanilla model.}
    \label{fig:ipw-baseline}
\end{figure}

%% file: sections/7_conclusion.tex
\section{Conclusion} \label{sec:conclusion}
We identify the need and promise of better understanding the unique behaviors of niche users in recommender systems. We examine users based not only on their item popularity preference but also on their activity level. We find that the quadrant of power-niche users is statistically significantly larger than expected in three benchmark datasets, whereas for mainstream users, the number of power users is below expectation. Further, using a proposed data valuation method, we show that power-niche user data is the most valuable for improving \emph{overall} recommendation performance, beyond being valuable for improving recommendations for niche users. From this motivating analysis, we develop \texttt{PAIR}, a reweighting framework that considers both user activity level and item popularity to reweight the BPR loss function. Instantiated on shallow and deep collaborative filtering models, \texttt{PAIR} upweights power-niche users and also Pareto-dominates the vanilla models. These results suggest that by considering user activity level, improving recommendations for niche users does not have to come at the cost of worsening recommendations for mainstream ones.

%% file: sections/appendix.tex
\appendix

\section{Notation}\label{sec:notation}
\input{sections/notation_table}

\section{Hyperparameter Optimization}\label{sec:optimized-hyperparam}
\input{sections/tables-and-figures/optimal-hyperparameters}

We optimize the hyperparameters for \texttt{PAIR} and all three baselines by first training on the training split and evaluating on the validation split. In the final evaluation, we train on the combined training and validation splits. We select the hyperparameters that yield the highest Recall@$k$ on the validation split. We perform the following grid searches: for \texttt{PAIR}, we search in $\alpha\in\{0, \frac{1}{4}, \frac{1}{2}, \frac{3}{4}, 1\} \times \beta \in \{0, \frac{1}{4}, \frac{1}{2}, 1\}$; for Pop. Reg. we search for $\gamma\in\{1, 10, 10^2, 10^3\}$; for IPW we search for $\beta\in\{0.1, \frac{1}{4}, \frac{1}{2}, 1\}$; for Pop. Comp. we search for $\alpha'\in\{10^{-2}, 10^{-3}, 10^{-4}, 10^{-5}, 10^{-6}\} \times \beta'\in\{0, \frac{1}{4}, \frac{1}{2}, \frac{3}{4}, 1\}$. The baseline method is applied to the Vanilla model, not the Vanilla-BPR one.

The optimized \texttt{PAIR} hyperparameters are listed in Table \ref{tab:optimal-hyperparams}. We utilize these parameters for the final evaluation in Table \ref{tab:results}. Observe that in five of the six configurations, $\alpha\in(0, 1)$, indicating that \texttt{PAIR} upweights based on user activity level more than the Vanilla model ($\alpha=0$) but not as much as the original BPR loss ($\alpha=1$). Our results show that explicitly tuning for the weight of activity level yields superior downstream performance and reduces popularity bias.

\section{Robustness Analysis of Quadrant Valuation Results}\label{sec:fixed-set-robustness}

Because the quadrant valuation results in Fig. \ref{fig:data-value} rely on a single fixed set $\mathcal{U}_{\text{fixed}}$, in this section, we analyze whether the value of power-niche user data is robust to alternative fixed sets. Recall from Section \ref{sec:data-value-methodology}, for a single fixed set, we calculate a quadrant value $\bar{\Delta}_q$, which is the average performance impact of treatment samples $\mathcal{S}(\mathcal{U}_q)$. In Fig. \ref{fig:data-value-fixed-set-robustness}, for each dataset and quadrant, we calculate $\bar{\Delta}_q$ for five fixed sets and report the average and $95\%$ confidence interval. Instead of sweeping over all treatment ratios, we set the treatment ratio to $0.3$. The figure shows that the value of power-niche users for both niche and overall performance is robust to alternative fixed set samples. In fact, Fig. \ref{fig:data-value-fixed-set-robustness} shows that after averaging over multiple fixed sets, the power-niche quadrant is the most valuable even for Yelp2018.
\input{sections/tables-and-figures/data-value-fixed-set-robustness}

\section{Supplemental Discussion on User-Weighting in BPR}\label{sec:BPR-sampling}

In this section, we provide further motivation for the user reweighting in \texttt{PAIR}. Recall that the BPR~\cite{rendle2009bpr} loss function reads,
\begin{equation}
    l^{\text{BPR}}(\mathbf{{Y}},\mathbf{\hat{Y}})
    =
    -\sum_{u=1}^n\sum_{i\in\mathcal{N}_u}\sum_{j\notin\mathcal{N}_u}
    \ln \sigma(\mathbf{\hat{Y}}_{ui}-\mathbf{\hat{Y}}_{uj}).
\end{equation}
As this form includes a sum over all triples of users, items they interacted with, and items that they didn't interact with, it can be costly to evaluate. This makes it natural to look for sample-based approximations of the BPR loss.
Replacing the empirical mean over the items in $\mathcal{N}_u$ and in $\mathcal{I}\setminus\mathcal{N}_u$ with the expectation under uniform sampling from these sets, the BPR loss can be rewritten as
\begin{equation}
    l^{\text{BPR}}(\mathbf{{Y}},\mathbf{\hat{Y}})
    =
    -\sum_{u=1}^n d_u(m-d_u)
    \mathbb{E}_{\substack{i\sim \mathrm{U}(\mathcal{N}_u) \\ j\sim \mathrm{U}(\mathcal{I}\setminus\mathcal{N}_u)}}\left[\ln\sigma(\mathbf{\hat{Y}}_{ui}-\mathbf{\hat{Y}}_{uj})\right],
\end{equation}
where $d_u=|\mathcal{N}_u|$ is the number of items a user $u$ interacts with and $m=|\mathcal{I}|$ is the total number items. Note that this is exact if $S_u=d_u(m-d_u)$ user-item-item triples per user are sampled without replacement.

As users usually only interact with a small subset of all items $m\gg d_u$, we will make the approximation $m-d_u\approx m$, and ignore this constant factor, which only results in a constant rescaling of the gradients during optimization.
The resulting loss is,
\begin{equation}
\begin{split}
    l^{\text{BPR}}(\mathbf{{Y}},\mathbf{\hat{Y}})
    &\approx
    -\sum_{u=1}^n d_u
    \mathbb{E}_{\substack{i\sim \mathrm{U}(\mathcal{N}_u) \\ j\sim \mathrm{U}(\mathcal{I}\setminus\mathcal{N}_u)}}\left[\ln\sigma(\mathbf{\hat{Y}}_{ui}-\mathbf{\hat{Y}}_{uj})\right]\\
    &=l^{\text{PAIR}}_{1,0}(\mathbf{{Y}},\mathbf{\hat{Y}}),
\end{split}
\end{equation}
where $l^{\text{PAIR}}_{1,0}$ is a particular instantiation of the \texttt{PAIR} loss $l^{\text{PAIR}}_{\alpha,\beta}$ defined in Eq.~\eqref{eqn:our-loss}. It is in this sense that we recover the original BPR loss for $\alpha=1,\,\beta=0$. Note, that \texttt{PAIR} samples triples of a user and two items \textit{with} replacement, using a user dependent sample size $S_u=\mathrm{round}\left(d_u^\alpha\frac{\sum_{u'\in\mathcal{U}}d_{u'}}{\sum_{u''\in\mathcal{U}}d_{u''}}\right)$.\\
LightGCN~\cite{he2020lightgcn} also employs a sample-based approximation of the BPR loss. However, in contrast to \texttt{PAIR}, LightGCN samples a constant number $S$ of user-item-item triples per user, independent of the number of items $d_u$ a user interacts with. This means LightGCN uses a loss of the form,
\begin{equation}
\begin{split}
    l^{\text{LGN}}(\mathbf{{Y}},\mathbf{\hat{Y}})
    &=
    -\sum_{u=1}^n S~
    \mathbb{E}_{\substack{i\sim \mathrm{U}(\mathcal{N}_u) \\ j\sim \mathrm{U}(\mathcal{I}\setminus\mathcal{N}_u)}}\left[\ln\sigma(\mathbf{\hat{Y}}_{ui}-\mathbf{\hat{Y}}_{uj})\right]\\
    &=l^{\text{PAIR}}_{0,0}(\mathbf{{Y}},\mathbf{\hat{Y}}),
\end{split}
\end{equation}
which is equal \texttt{PAIR} for the hyperparameter setting $\alpha=0,\,\beta=0$.

%% file: sections/notation_table.tex
\begin{table}[h]
    \centering
    \caption{Notations used in this paper.}
    \begin{tabular}{p{0.18\linewidth} p{0.77\linewidth}}
        \toprule
        \textbf{Symbol} & \textbf{Meaning} \\
        \midrule
        $\mathcal{U}, \mathcal{I}$ & Set of users and items \\
        $n, m$ & Number of users and items \\
        $\mathbf{D}$ & $n\times m$ binary user–item interaction matrix \\
        $\mathbf{X}, \mathbf{Y}$ & Training and test splits of $\mathbf{D}$ \\
        $\mathbf{Y}_{ui}, \hat{\mathbf{Y}}_{ui}$ & Ground-truth and predicted score for $u,i$ \\
        \midrule
        $\mathcal{N}_u$ & Set of items user $u$ has interacted with \\
        $d_u, d_i$ & Activity level of user $u$ and popularity of item $i$\\
        \midrule
        $R$ & An evaluation function e.g. Recall@$k$\\
        $\textrm{U}(x)$ & Uniform distribution over a set $x$ \\
        $\mathcal{S}(x)$ & A uniform sample (with replacement) over $x$ \\
        $S_u$ & Number of triplets sampled for user $u$ \\
        $p_i$ & Propensity of item $i$ (used in IPW baseline) \\
        $\mathcal{U}_{\text{fixed}}, \mathcal{U}_q$ & Fixed user set and treatment quadrant user set \\
        \midrule
        $\mathbf{u}_u, \mathbf{v}_i \in \mathbb{R}^d$ & $d$-dimensional embeddings for user $u$ and item $i$ \\
        $l(\mathbf{Y}, \hat{\mathbf{Y}}), \sigma(x)$ & Loss function and sigmoid function \\
        $\alpha \in [0,1]$ & Hyperparam. for user activity weighting \\
        $\beta \in [0,\infty)$ & Hyperparam. for item popularity weighting \\
        $\gamma$ & Regularization strength (Pop. Reg. baseline) \\
        $\alpha', \beta'$ & Hyperparams for popularity-compensation\\
        \bottomrule
    \end{tabular}
    \label{tab:notation}
\end{table}

%% file: sections/tables-and-figures/optimal-hyperparameters.tex
\begin{table}[h]
\centering
\caption{Optimal hyperparameters $(\alpha, \beta)$ for \texttt{PAIR} by model and dataset.}
\label{tab:optimal-hyperparams}
\begin{tabular}{l l c c}
\toprule
\textbf{Model} & \textbf{Dataset} & $\boldsymbol{\alpha}$ & $\boldsymbol{\beta}$ \\
\midrule
LGN & Gowalla     & 0.5  &  0.0  \\
LGN & Yelp2018    & 1.0  & 0.25 \\
LGN & Amazon-Book & 0.25 & 0.25 \\
MF  & Gowalla     & 0.0  & 0.5  \\
MF  & Yelp2018    & 0.5  & 0.5  \\
MF  & Amazon-Book & 0.75 & 0.0  \\
\bottomrule
\end{tabular}
\end{table}

%% file: sections/tables-and-figures/data-value-fixed-set-robustness.tex
\begin{figure}[ht]
    \centering
    \begin{subfigure}{\columnwidth}
        \centering
        \includegraphics[width=\columnwidth]{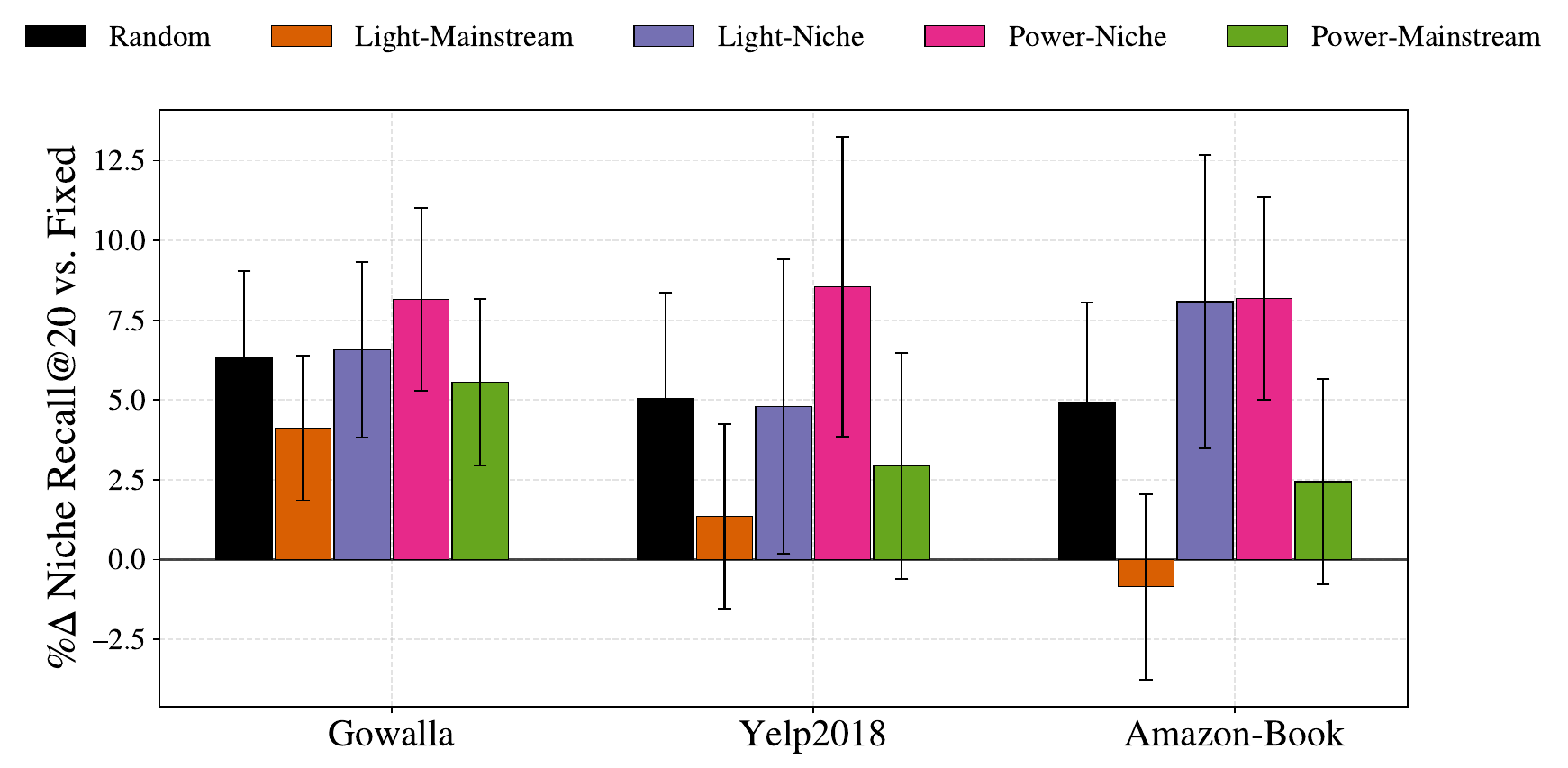}
    \end{subfigure}
    \begin{subfigure}{\columnwidth}
        \centering
        \includegraphics[width=\columnwidth]{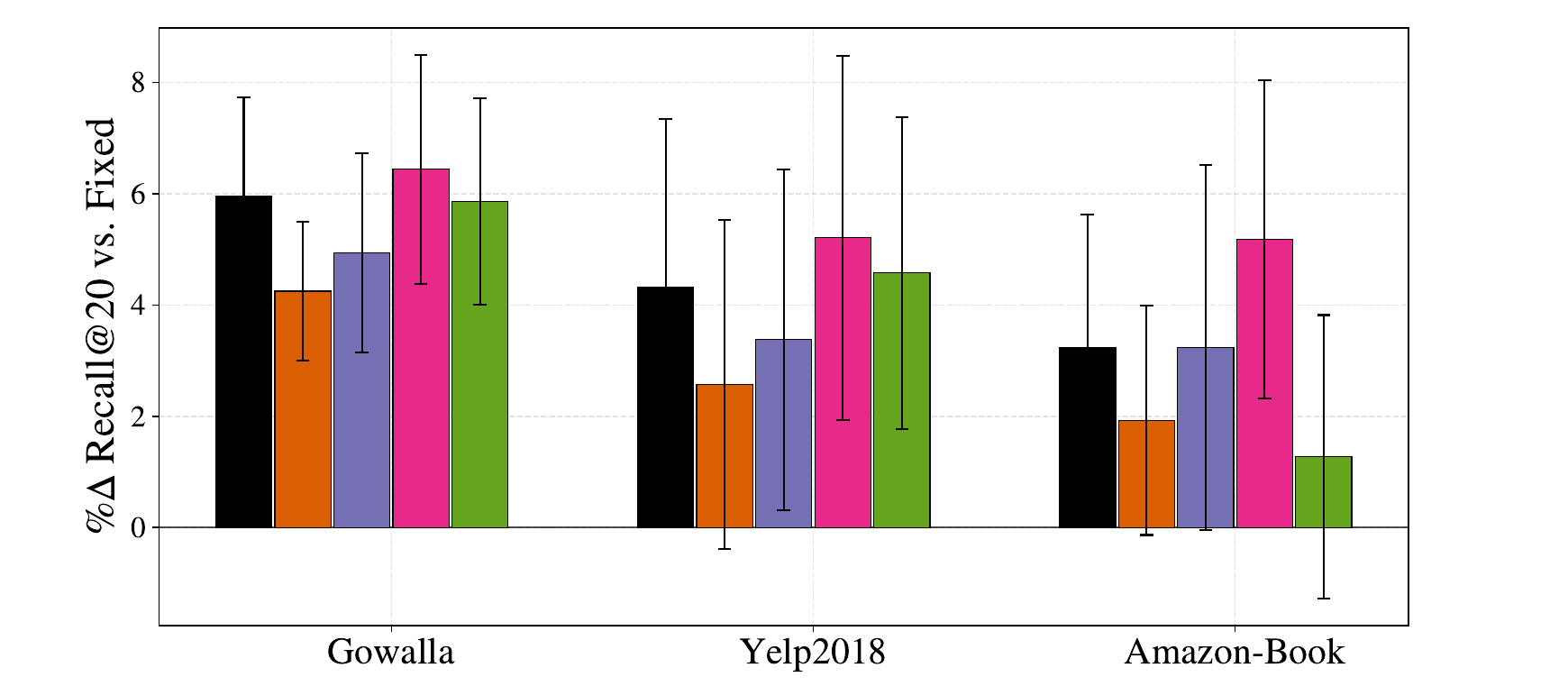}
    \end{subfigure}
    \Description{Grouped bar chart showing the relative change in recommendation performance when training data from different user quadrants are added under multiple alternative fixed user sets. Bars correspond to user quadrants and are grouped by dataset and evaluation metric. Across all fixed-set choices, adding power-niche users consistently yields the largest improvements in niche and overall performance, indicating robustness of the valuation results to the choice of fixed users.}
    \caption{Value of each quadrant averaged over multiple fixed sets. While the quadrant valuation results in Fig. \ref{fig:data-value} use a single fixed set, in this figure, we average the quadrant values across five fixed sets. The treatment ratio is set to $0.3$. Error bars indicate $95\%$ confidence intervals. The figure shows that the value of power-niche users for niche and overall performance is robust to alternative fixed-set samples.}
    \label{fig:data-value-fixed-set-robustness}
\end{figure}